\documentclass{rspublic}
\pdfoutput=1
\usepackage{amsmath, amssymb}
\usepackage[pdftex]{graphicx}
\usepackage{ifpdf}

\newcommand{\mathsym}[1]{{}}

\usepackage{natbib}

\newif\iflatextortf
\iflatextortf

\fi

\begin{document}

\title[Statistical mechanics of evolving traits]{The statistical mechanics of a polygenic characterunder stabilizing selection, mutation and drift}

\author[H.P. de Vladar and N.H. Barton]{Harold P. de Vladar$^1$ and Nick H. Barton$^2$}

\affiliation{$^1$Institute of Science and Technology Austria (IST Austria) \\ Am Campus 1,Klosterneuburg A-3400, Austria \\  $^2$Institute of Evolutionary Biology, School of Biological Sciences, University of Edinburgh, Kings Buildings, Edinburgh EH9 3JT, UK}

\label{firstpage}

\maketitle

\begin{abstract}{ Evolution; Quantitative Genetics; Statistical Mechanics; Entropy; Epistasis; Genetic Variance, Shifting Balance.}
By exploiting an analogy between population genetics and statistical mechanics, we study the evolution of a polygenic trait under stabilizing selection, mutation, and genetic drift. This requires us to track only four macroscopic variables, instead of the distribution of all the allele frequencies that influence the trait. These macroscopic variables are the expectations of: the trait mean and its square, the genetic variance, and of a measure of heterozygosity, and are derived from a generating function that is in turn derived by maximizing an entropy measure. These four macroscopics are enough to accurately describe the dynamics of the trait mean and of its genetic variance (and in principle of any other quantity). Unlike previous approaches that were based on an infinite series of moments or cumulants, which had to be truncated arbitrarily, our calculations provide a well-defined approximation procedure. We apply the framework to abrupt and gradual changes in the optimum, as well as to changes in the strength of stabilizing selection. Our approximations are surprisingly accurate, even for systems with as few as 5 loci. We find that when the effects of drift are included, the expected genetic variance is hardly altered by directional selection, even though it fluctuates in any particular instance. We also find hysteresis, showing that even after averaging over the microscopic variables, the macroscopic trajectories retain a memory of the underlying genetic states.
\end{abstract}

\section{INTRODUCTION}

Quantitative genetics aims to explain and describe the response of a heritable trait to selection, but in the classical approach, dispensing with
the genetic details that account for that heritability (Lande 1979; Turelli 1988; Lynch and Walsh 1998). Yet, population genetics tells us that such
responses cannot be predicted without knowing those genetic details (Turelli 1988; Barton and Turelli 1987; B{\" u}rger 2000). Paradoxically, we
can indeed make such predictions from empirical measurements of traits, but only by assuming that the genetic variance (the variance of the trait
caused by genetic differences) is fixed: changes in variance due to selection have been hard to predict (e.g. Hill 1982; Zhang and Hill 2010).

Directional selection (DS) will usually allow a quick response at the cost of the depletion of genetic variance, which will ultimately be due to
mutation around a unique optimal genotype ( B{\" u}rger 2000, ch IV). However, most traits are thought to be under some kind of stabilizing selection
(SS), in which case the genetic variance will be reduced by fixing any one of the many genotypes that match the optimal value (Wright 1935a; Bulmer1972;
Barton 1986). In order to understand the full stochastic model, that is a polygenic trait under SS, mutation and drift --the goal of
this article-- we first summarize the theory for infinite populations  (Wright 1935a 1935b; Barton 1986), where random fluctuations are absent,
and evolution is deterministic.

For simplicity, we assume that each gene has only two alternative states (termed \textit{alleles}). When mutation ($\mu $) is much weaker than the
selection ($\mathit{s}$) on each allele, the trait mean converges to the optimum, because any genetic locus can be almost fixed for either allele
(i.e., allele frequencies are close to 0 or 1). Thus, there will be many stable equilibria for the whole set of loci, at which any genotype that
matches the optimum is near fixation - there are degenerate genetic states for a trait mean matching the optimum. Assuming that there are enough
genes ($n>\!>$1), the genetic variance is then roughly $4n\mu /\mathit{s}$ (Barton and Turelli 1987). When this variance is
much greater than the maximum that can be contributed by any one locus (Barton 1986), its increase due to a substitution at any one locus is much
smaller than the existing variance, as we expect to be the case for a trait that is influenced by many genes. In this regime, there are many sub-optimal
stable equilibria, for which the genotype that is close to fixation does not quite match the optimum. For example, if the optimum is at zero, then
with 100 loci, there are \(\frac{100!}{(50!)^2}\sim 10^{29}\) equilibria that match this precisely, but asymmetrical equilibria with 49 $\texttt{'}$-$\texttt{'}$
and 51 $\texttt{'}$+$\texttt{'}$ loci may also be stable. However, selection adjusts the frequencies of rare alleles with these asymmetrical combinations,
such that the mean is still very close to the optimum, but the genetic variance is substantially higher, and the mean fitness is lower as a consequence
(Barton 1986).

\begin{figure}[t]
\begin{center}
\includegraphics[scale=0.65]{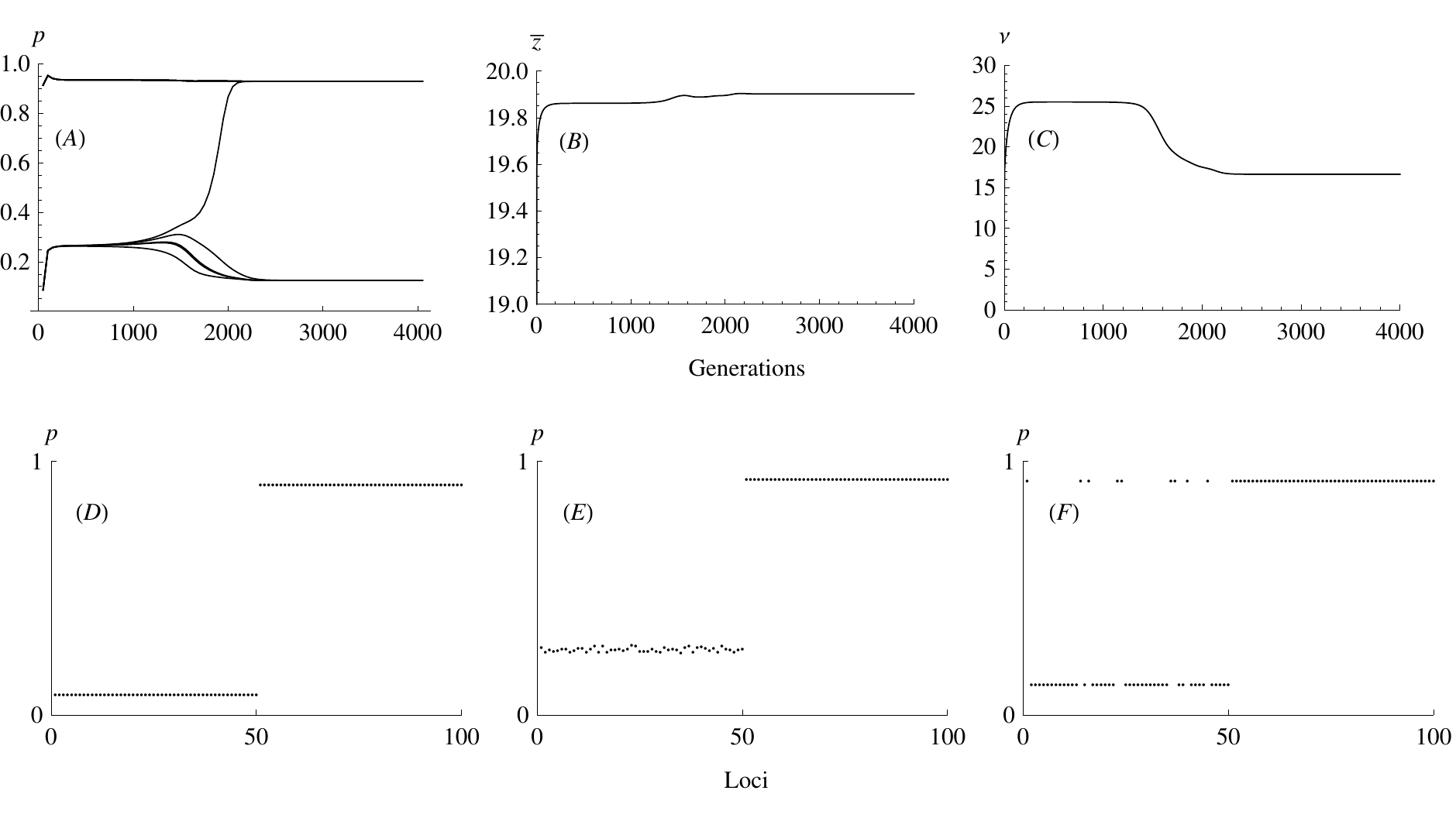}
\caption{
{
Top row: (A) Allele frequencies, (B) trait mean, and (C) genetic variance plotted against time.  A population
is initially at equilibrium with stabilizing selection \(s=0.05\) towards \(z_{\text{opt}}=0\) acting on an additive trait, with \(n=100\) loci of
effect $\gamma $=1; the mutation rate is $\mu $=0.002 per locus, which maintains a genetic variance of \(\nu  =4n \mu /\mathit{s} =16\).  The optimum
then shifts abruptly to \(z_{\text{opt}}=20\), and the mean responds almost immediately (top, middle).  The variance increases abruptly (top right)
as ,the allele frequencies at all the $\texttt{'}$-$\texttt{'}$ loci increase substantially (A, lower left region).  However, this new state is
unstable, and slight variations in the initial conditions cause some loci to shift down, and some to shift up.  As a result, the genetic variance
returns to its original value.  The lower row shows snapshots of allele frequencies at times (D) 0, (E) 800 and (F) 3000 generations.}
}
\label{Fig:DeterministcExample}
\end{center}
\end{figure}

Imagine that the population is initially at an $\texttt{'}$optimal$\texttt{'}$ equilibrium, which maximizes mean fitness. If the optimum now shifts,
the population will stay near fixation for the original genotype, and so the genetic variance will inflate as allele frequencies adjust to keep the
mean near the optimum. At some point, the equilibrium becomes unstable, one or more loci substitute, and the genetic variance drops abruptly. If
all alleles have the same effect, as assumed so far, then variance will inflate at many loci, and so may become very high (fig. \ref{Fig:DeterministcExample}). However, with
unequal effects, the loci with smallest effect will substitute first, and the overall fluctuation will be much smaller.

If the trait mean lags far behind the current optimum, we would expect to see erratic trajectories in an infinite population because of the following
two points:
\begin{enumerate}
\item[(i)] very rare alleles will increase, and will sweep through at an unpredictable time that depends on their initial frequency.

\item[(ii)] The various --and phenotypically equivalent-- combinations of alleles are reorganized when the phenotypic optimum
takes a new value, transiently inflating the variance, as explained above. 
\end{enumerate}

We envisage three factors that smooth potentially erratic phenotypic paths. First, if the effects of alleles vary across loci, there is the opportunity
to make smaller steps as the optimum changes, which alleviates pronounced and coincident changes in allele frequencies, and diminishes the excess
variance as the phenotypic optimum moves. Second, if we average over all the possible genetic combinations that match the phenotypic optimum, then
we expect that the speed and direction of the response to a change in optimum will be more regular. Third, if the population is finite, random drift
will on the one hand homogenize the initial distribution of allele frequencies, alleviating point \textit{i} above, and on the other, facilitate
transitions between $\texttt{'}$adaptive peaks$\texttt{'}$, alleviating point \textit{ii}, so that the population can evolve without being trapped
at local equilibria, as in Wright's (1931b) $\texttt{'}$shifting balance$\texttt{'}$. Also, note that under drift, plus weak mutation,
allele frequencies $p$ are distributed as \(\sim [p  (1-p)]^{-1}\), a special case in which the genetic variance stays constant under DS
(since with this distribution, the increased variance due to rare alleles that become common is baalnced by the loss of variance from common alleles
that become rare). Here, rare alleles replenish variance at the same rate as common alleles are fixed.

Measurable quantitative genetic variables are not sufficient to predict response to selection: the dynamics of the trait mean and of the genetic
variance under selection depend on allele frequencies. We can use moments instead, but their dynamics depend on higher moments, and we end up requiring
as many variables as allele frequencies (Barton and Turelli 1987; B{\" u}rger 1991 1993 2000; Turelli and Barton 1994). In part, this problem arises
from \textit{i} and \textit{ii} above. Surprisingly, including a third evolutionary process, random genetic drift, allows us to employ a different
mathematical toolbox (from physics' statistical mechanics, SM) to make reliable predictions. Although we can no longer predict a single trajectory
in detail, we can instead predict expectations over the ensemble of trajectories. Thus, we will focus on the second and third factors mentioned above,
namely the averaging over allele frequencies that is caused by random genetic drift.

\subsection{Statistical Mechanics}

The need to specify the detailed genetic state is avoided by the development of an analogy between SM and population genetics (Pr{\" u}gel-Bennet and Shapiro 1994; 1997; Rattray 1995; Pr{\" u}gel-Bennet 1997; Rogers and Pr{\"u}gel-Bennett 2000). In these works, information entropy (Shannon, 1948) was maximized under the assumption that any initial state was a priori equally likely. However, this choice of prior is not based on population genetics: we know that states are not equiprobable under random genetic drift, because in a finite population, fluctuations bias the neutral distribution towards the boundaries. Further work developed this idea in detail, tailoring an entropy measure that ensures that the equilibrium distribution matches the stationary solution to the diffusion equation (Barton and de Vladar 2009, Barton and Coe 2009). This allows us to collapse an arbitrary number of degrees of freedom that cannot be accurately known (e.g. allele or genotype frequencies, central moments, cumulants, etc.) into a few variables, which approximate the whole stochastic system. The theory is built in two stages.

\noindent \pmb{\textup{ Maximum entropy.}} First, we observe that a measure of \textit{entropy} is maximized at equilibrium, subject to constraints on the expectations of the macroscopic variables (Barton 1989; Aita et al 2004 2005; Sella and Hirsch 2005; Barton and de Vladar 2009). (Alternatively, we can define a $\texttt{'}$free fitness$\texttt{'}$ as the entropy divided by population size, plus the mean fitness, which is maximized at equilibrium under selection and drift; this is analogous to free energy in thermodynamics (Iwasa 1988, Barton and Coe 2009)). Although entropy is a concept alien to population genetics, it summarizes two fundamental aspects of the equilibrium distribution. On the one hand, it measures the degree of divergence of the distribution of the allele frequencies from a neutral base distribution. On the other hand, the entropy couples two sets of variables, the \textit{microscopic} that in this case we take to be the allele frequencies, and the \textit{macroscopic} that are the expectations of quantitative variables.

Generally speaking, the SM approach takes advantage of genetic drift in two ways. To start with, drift is expected to spread the trajectories relaxing the problem of trapping at local fitness optima, just as envisaged by Wright (1932). To follow, instead of calculating the values of metric characters themselves -- these are stochastic variables-- it makes predictions for their expectations. 

Although we are free to apply any kind of selection, it is crucial that it act only on the chosen macroscopic variables. That is, we cannot assume that selection will act on specific genotypes, because then, selection could favour arbitrary and improbable macroscopic states: the maximum entropy Ansatz would be contradicted, and would inevitably fail to give accurate predictions. For example, if we force the population to a specific set of allele frequencies, \(p_i\), then under directional selection \(s\) there would be sudden jumps at times $\sim $ \((1/s)\log \left(1\left/p_i\right.\right)\), as alleles reach high frequency.

The distribution of allele frequencies --like any statistical distribution-- is formally a function of a set of parameters. In population genetics, it typically depends on the selection coefficients, mutation rate, and population size, and the distribution of macroscopics can be calculated from these, using the diffusion approximation. This averaging over the microscopic variables has the fortunate consequence that the distribution of macroscopics, as well as their rates of evolution, do not explicitly depend on the allele frequencies, although their distribution is in fact estimated implicitly by maximizing the entropy. This averaging is not straightforward under stabilizing selection, because there is epistasis for fitness, which fully couples the distribution of allele frequencies. However, the calculations can be greatly simplified because under the central limit theorem, the distribution of the macroscopic variables, conditioned on a particular value of the trait mean, will be approximately Gaussian. This will allow us to eliminate any explicit dependence on the allele frequencies, and gives workable expressions for the expectations.

\noindent \pmb{Quasi-equilibrium}. Second, we make a quasi-equilibrium approximation, which assumes that even during periods of change, the distribution of allele frequencies adopts a shape that takes the form of an equilibrium distribution, whose parameters are chosen so that the distribution matches the current macroscopic variables.

In this framework, we can describe the state of the population by two complementary sets of variables: the expectations of the chosen macroscopics \(\left(\left\langle \bar{z}\right\rangle , \langle v\rangle  \ldots \right)\), and the corresponding selective forces on these macroscopics.  This is precisely analogous the complementary set of variables (forces and variables of state) in statistical mechanics.  Rather than working with the rates of evolution of the macroscopic variables, we can calculate the rates of change of the evolutionary forces. Consequently, we can find the expectations of the quantitative variables at any time, and thus know their evolutionary trajectory. If parameters change sufficiently slowly, the ensemble will be close to an equilibrium distribution, as with reversible changes to a physical system. When parameters change quickly, however, nothing ensures that the distribution will be close to some stationary distribution. Thus, the present work aims in part to verify this assumption, by comparing the results with numerical simulations. The mathematical details of this method are explained in detail by Barton and de Vladar (2009).

A crucial advantage of the SM method is that it avoids the recursion to higher moments (or cumulants) of the traits, where higher order terms must be neglected arbitrarily (Barton and Turelli 1987 1990; B{\" u}rger 1991 1993 2000; Turelli and Barton 1994; Rattray and Shapiro 2001): the key is the choice of an appropriate set of macroscopics, which is determined by the set of traits hat determine fitness.

Previously, we have successfully applied these ideas to the much simpler case of DS on a polygenic trait subject to mutation and drift (Barton and de Vladar 2009). There, our predictions are accurate for many loci of arbitrary effects. However, under DS solving the problem for a trait with $n$ loci amounts essentially to solving the problem for a single locus, since there is no coupling between loci. Thus it is to some extent, a trivial case. - though not entirely so, because maximum entropy still drastically collapses the degrees of freedom from the full allele frequency distribution to only two, the trait mean, and the log-heterozygosity (eq. \ref{Eq:GeneticVariability}). A more demanding situation, and of deep biologically relevance, is SS, which is the main focus of our present work. We will study whether the SM method applies to the challenging case of SS, mutation, and drift. In particular, we will develop the details using Gaussian SS as a model. We analyze sharp changes in the optimum, a steadily moving optimum, and changes in selection on the genetic variance.

\section{STABILIZING SELECTION, MUTATION AND GENETIC DRIFT}

\subsection{Distribution of allele frequencies}

We assume throughout that recombination is fast relative to selection, mutation and drift, so that we can assume linkage equilibrium, and describe the population by its allele frequencies. We also assume that populations size is not small, so that we can use the diffusion approximation. The fitness of an individual with trait $z$ is given by \(\log \left[W_z\right] = -\frac{\mathit{s}}{2}\left(z-z_o\right) ^2\), which penalizes deviations from the optimum \(z_o\), since the extremes are unfit. Notice that this expression can be written as \( -\frac{\mathit{s}}{2}\left(z-\bar{z}\right)^2-\frac{\mathit{s}}{2}\left(\bar{z}-z_o\right) ^2 +\)\(2\frac{\mathit{s}}{2}\left(z-\bar{z}\right)\left(\bar{z}-z_{\text{opt}}\right)\). The first term suggests that individuals that deviate from the mean of the population have less fitness. The second term is the deviation of the trait mean from the optimum. The mean log fitness is thus \(\overline{\log [W]} \sim  \log \left[\bar{W}\right]\simeq -\frac{\mathit{s}}{2}\nu -\frac{\mathit{s}}{2}\left(\bar{z}-z_{\text{opt}}\right) ^2 \). Here we neglect terms \(\mathcal{O}\left(\mathit{s}^2\right)\), and the third term above averages to zero. For further convenience, we expand the quadratic term, and express it as \(\log \left[\bar{W}\right]=\sigma \nu +\beta \bar{z}-\lambda \bar{z}^2 \)+ const. This is essentially the same, but where selection is acting $\texttt{'}$independently$\texttt{'}$ over three quantitative variables: selection (of strength $\sigma $) against the genetic variance, $\nu $; a directional term of strength $\beta $ (\(=-z_{\text{opt}} \mathit{s}\)) regressing the mean trait \(\bar{z}\) to the optimum, and selection against deviations from the actual mean (of strength \(\lambda \)). Most works consider that \(\sigma =\lambda =-\mathit{s}/2\), but here this constraint will be relaxed and allow $\sigma $ and $\lambda $ to differ, in order to improve the quasi equilibrium approximations, as will become clear later. We can think of examples, such as sexual selection or resource exploitation, where the individuals that deviate too much from the population's current mean value are unlikely to leave offspring, or selection might act directly on heterozygotes (and hence directly on $\nu $). Of course, in some of these cases, selection might be frequency-dependent, in which case we will not have a stationary distribution that is of potential form, and the following analyses might not hold. However, we will keep these issues aside and concentrate on the simpler case of frequency independent selection, but allowing selection on the variance and on the deviation of the trait mean (factors $\sigma $ and $\lambda $, respectively) to differ.\\
We will assume that selection is weak relative to recombination, so that linkage equilibrium holds. Then, if the population is moderately large, the evolution of the frequency \(p_l\) of a beneficial allele at each locus can be described by
\begin{equation}
\label{Eq:Wright-FisherModel}
\frac{\rd  p}{\rd  t}=\frac{p  q}{2}\frac{\partial \log\left[\bar{W}\right]}{\partial p}-\mu (2p-1)+\zeta _d
\end{equation}
where $\mu $ is the mutation rate (assumed to be reversible and equal in both directions), and \(\zeta _d\) is the stochastic term due to drift, modelled as a Gaussian of mean zero and variance $pq/2N$ where $q=1-p$, and $N$ is the effective size of the population. Notice that \(\bar{W}\) depends on \(\bar{z}^2\) and so couples the dynamics at all loci, something that does not occur under DS. In other words, fitness is essentially epistatic.

At equilibrium, this process has the joint distribution of allele frequencies, \(\psi (\pmb{p}) \) (Wright 1931; Kimura 1965; Crow and Kimura 1970, pp. 442):
\begin{equation}
\label{Eq:DistributionAlleleFrequencies}
\psi (\pmb{p})=\frac{\exp\left[2N \beta \bar{z}-2N\!\lambda \bar{z}^2+2N\!\sigma \nu +2N\!\mu U\right]}{\mathbb{Z} \prod _{i=1}^n p_iq_i}
\end{equation}
which is identical to the more familiar expression \(\psi (\pmb{p})=c \bar{W}^{2N}\prod _{i=1}^n \left(p_iq_i\right) ^{4N\!\mu-1}\) but arranged in a different way. The quantity $\mathbb{Z}$ normalizes the distribution. (From now on, we will use bold characters to denote vectors, and script capitals to denote matrices). As explained before (Barton 1989; Barton and de Vladar 2009; Barton and Coe 2009), we describe the effects of mutation by a potential function $U$, which is essentially the log of the geometric mean heterozygosity:
\begin{equation}
\label{Eq:GeneticVariability}
U =2 \sum_{i=1}^n \log[p_i q_i]
\end{equation}

Thus we have a set of macroscopic variables \(\pmb{A}=\left\{\bar{z},\bar{z}^2, \nu , U\right\}\) that is associated with the parameters (forces) \pmb{ $\alpha $}=$\{\beta $, $\lambda $, $\sigma $, $\mu \}$. Both together completely determine the distribution of allele frequencies. At first sight it might seem that specifying both \(\bar{z}\) and \(\bar{z}^2\) is redundant. However, the latter introduces epistasis to the fitness function, making it non-multiplicative. But also, since it is their expectations that constrain the entropy, both terms are necessary to couple the microscopic and macroscopic variables,  and thus to determine the expectations of the mean and the variance of the trait.

\subsection{Additive polygenic traits}

So far, we made no assumptions about the trait. The set of variables just defined, \pmb{ \textit{A}}, depend arbitrarily on the allele frequencies. We will assume now some elementary properties. Each trait is composed of a pair of alleles (thus we are assuming diploids) that contribute additively, each with an effect \(\gamma _l\). The labels \textit{X} denote different alleles (\textit{X}=0,1). Under this model, the trait, its mean, and its variance in the population are respectively
\begin{eqnarray}
z = \sum _{i=1}^n \gamma _i\left(X_i+X_i'-1\right)\\
\bar{z} = \sum _{i=1}^n \gamma _i\left(2p_i-1\right)\\
\nu  = 2\sum _{i=1}^n \gamma _i^2p_iq_i
\end{eqnarray}

Defined this way, the trait is in the range of \(\pm \sum _l \gamma _l\), and the genetic variance is in \(\left[0,\nu_{\max }\right.\)], with \(\nu_{\max }=\frac{1}{2}\sum _{l=1}^n \gamma _l ^2\). If we assume that all loci have equal effects $\gamma $, then \(\bar{z}=\gamma  \sum _{l=1}^n \left(2p_l-1\right)\), and \(\nu =2 \gamma ^2 \sum _{l=1}^n p_lq_l\). Without loss of generality, we will assume $\gamma $=1. We can always include \(\gamma \neq 1\) by scaling \(\beta \to \gamma \beta \) and \(\sigma \to \gamma ^2\sigma \).

Along this article, we assume that mutation ($\mu $) is weaker than selection ($\mathit{s}$) on each allele, that is \(\mu <\mathit{s}\gamma ^2/4\). But at the same time, since we require that the standing genetic variance is grater than the variance due to a substitution, we will work in the regime of  \(\mathit{s} \gamma ^2/4<n \mu \) (Barton 1986), which ensures that there are many stable genetic equilibria.

\section{STATISTICAL MECHANICS}

In our previous work, we explained in detail how the distribution in eq. (\ref{Eq:DistributionAlleleFrequencies}) can be derived by maximizing a particular entropy measure (Barton and de Vladar 2009). In population genetics, this uncertainty measures the possible states in which a population can be. It is a functional of the distribution of allele frequencies $\psi $(\pmb{ \textit{p}}). We need to employ an information entropy measure which is defined relative to the distribution under drift alone (Barton and de Vladar 2009; Barton and Coe 2009). That is, in the absence of any other evolutionary factor, random drift would drive the evolution of the population towards the distribution\( \phi =\prod _{l=1}^n \left(p_lq_l\right) ^{-1}\), which is a U-shaped distribution (Wright 1931). In fact, $\phi $ is not properly a distribution because it cannot be normalized. Nevertheless, under drift alone $\phi $ is the solution to the diffusion equation (Kimura 1962), and we will set entropy so that with a base distribution $\phi $, it gives the correct stationary distribution with mutation and selection. Accordingly, we defined the entropy as
\begin{equation}
S(\psi ) = -\frac{1}{2N}\int \psi  \log(\psi /\phi ) d^n\pmb{p}
\end{equation}

When \textit{S }is maximized with respect to $\psi $ we get that \(\psi \propto  \phi \). Although $\phi $ will diverge when \textit{pq}$\to $0, imposing constraints on \textit{S} will ensure that this is not a problem. This particular entropy measure was postulated to ensure that the distribution of allele frequencies corresponds to the one obtained by the diffusion methods, i.e. eq. (\ref{Eq:DistributionAlleleFrequencies}). However, the probability distribution is changed in the presence of, say, mutation and/or selection. Thus, entropy changes under any other factor that affects the evolution of a character or of fitness, like selection or mutation, because these increase or decrease the uncertainty of the outcomes, biasing the distribution $\psi $. To include these effects, we require that the distribution of allele frequencies must have specific values for the expectations $\langle $\pmb{ \textit{A}}$\rangle $. This is introduced in the form of constraints to the entropy maximization problem, and thus we introduce a Lagrange multiplier \(\alpha _k\) for each constraint \(\left\langle A_k\right\rangle  = \int A_k \psi  d^n\pmb{p} \). Thus, we need to maximize the following functional (on $\psi $):
\[-\frac{1}{2N}\int \psi  \log(\psi /\phi ) d^n\pmb{p} + \sum _j \alpha _j\left\langle A_j\right\rangle +\alpha _0\int \psi  d^n\pmb{p} .\]
Here the third term ensures normalization of the distribution. This leads exactly to the distribution of allele frequencies in eq. (\ref{Eq:DistributionAlleleFrequencies}), and thus the Lagrange multipliers are the forces (selective coefficients, and mutation rate). The normalization condition is:
\begin{equation}
\label{Eq:DefPartitionFunction}
\mathbb{Z} = \int\exp\left[\sum _j 2N\!\alpha_jA_j\right] \phi (\pmb{p})d^n\pmb{p} ,
\end{equation}
where we renamed the multiplier \(\exp\left[-2N  \alpha _0\right.\)]$\equiv \mathbb{Z}$. This quantity is a function of the parameters  \(\alpha _j, (j\neq 0)\) and it is analogous to the \textit{partition function} in statistical physics. It plays a fundamental role, because it allows direct calculation of the macroscopics, as well as their covariances, since Log[$\mathbb{Z}$] is a generating function: \(\partial _{\pmb{\alpha }}\log [\mathbb{Z}]=2N\langle \pmb{A}\rangle \), and \(\partial ^2_{\pmb{\alpha ,\alpha }}\log[\mathbb{Z}]=4N^2\mathcal{C} \), where \(\mathcal{C}=\text{cov}\left(A_i,A_j\right)\) is the covariance matrix of the macroscopics. These expectations and covariances are with respect to the distribution $\psi $, that is across possible states of a population, or alternatively with respect to an ensemble of populations, and must not be confused with the means and variances of a trait within a population. The technical effort focuses on a workable calculation for $\mathbb{Z}$, which will lead to expressions for the rates of change of the \pmb{ $\alpha $} and the \(\langle \pmb{A}\rangle \). In general, any vector of quantitative variables, in our case \(\pmb{A}\), that is a function of the allele frequencies will have an expectation (macroscopic) that evolves as
\begin{equation}  \label{Eq:GlobalRateOfChange}
\frac{\rd}{\rd  t}\langle  \pmb{A} \rangle  =\mathcal{B}.\pmb{\alpha } +\pmb{V}
\end{equation}
where $\mathcal{B}$ is a generalized matrix of genetic covariances, and \pmb{ \textit{V}} the rate of random drift; \(\mathcal{B}_{i  j} = \left\langle  \sum _{l=1}^n \frac{\partial A_i }{\partial p_l}\frac{ p_lq_l}{2}\frac{\partial A_j }{\partial p_l}\right\rangle \) and \(V_i = \left\langle \sum _{l=1}^n  \frac{p_lq_l}{4N}\frac{\partial ^2A_i }{\partial p_l^2}\right\rangle  \). These formulas can be derived straightforwardly in several ways. The most compelling for a biologist is to use Wright's (1937) formula, via the diffusion equation (Ewens 1979, pp. 136-137), or from the Fokker-Planck equation (for eq. \ref{Eq:Wright-FisherModel}; Gardiner 2004, ch. 5). Neither \pmb{ \textit{V}}, $\mathcal{B}$, nor $\mathcal{C}$ (defined above) assume additive or equal effects. However to obtain specific expressions we need to make some assumptions. We will now assume additivity, for which we get
\begin{equation} 
\label{Eq:ExpectationRateMatrix}
\mathcal{B} = \left\langle 
\begin{array}{cccc}
 \nu  & 2\bar{z}\nu  & m_3 & -2\bar{z} \\
 2\bar{z}\nu  & 4\bar{z}^2\nu  & 2\bar{z}m_3 & -4\bar{z}^2 \\
 m_3 & 2\bar{z}m_3 & m_4 & 4\left(\nu _{\max }-\nu \right) \\
 -2\bar{z} & -4\bar{z}^2 & 4\left(\nu _{\max }-\nu \right) & H
\end{array}
\right\rangle ,
\end{equation}
and
 \begin{equation}
\pmb{V} = \left\langle 0,\frac{\nu }{N},-\frac{\nu }{2N},-\frac{H+4n}{2N}\right\rangle
\end{equation}

These expressions are valid also under unequal effects. (The unfamiliar macroscopics involved, namely \(m_3, m_4,\) and \textit{H} are specified in Appendix 1). Note that the SM method provides a way of directly calculating the expectations of these quantities, as well as of the higher moments of the trait: they follow from mathematical manipulations of the generating function $\mathbb{Z}$ (Barton and de Vladar 2009). This is one of the most important improvements from the SM method with respect to other moments or cumulant methods where these higher order moments have to be computed \textit{ab initio} using their particular differential equations, and approximated using specific models, like the Gaussian or House of Cards models (Barton and Turelli 1989), perturbation analysis (Rattray and Shapiro 2001), and/or simply by truncating the equations at a convenient level (Barton and Turelli 1987; B{\" u}rger 1991 1993 2000; Turelli and Barton 1994; Rogers and Pr{\" u}gel-Bennett 2000). \\
Under the quasi-equilibrium approximation, we assume that at every time point there is a set of variables \(\pmb{\alpha }_{(t)}^*\) (for short, \(\pmb{\alpha }^*\)) which would give the observed $\langle $\pmb{ \textit{A}}$\rangle $ at a steady state, so that \(\frac{\rd \langle \pmb{A}\rangle }{\rd  t}=0\); then \(\pmb{V}^*=-\mathcal{B}^*\cdot \pmb{\alpha }^*\). Thus, the matrix $\mathcal{B}$ in eq. (\ref{Eq:GlobalRateOfChange}) is approximated by its equilibrium expression, but evaluated at \(\pmb{\alpha }^*\), denoted \(\mathcal{B}^*\). Then, substituting into eq. (\ref{Eq:GlobalRateOfChange}), we get that \(\frac{\rd\langle \pmb{A}\rangle }{\rd  t} =\mathcal{B}^*\cdot \left(\pmb{\alpha }-\pmb{\alpha }_{(t)}^*\right)\). Since the matrix \(\mathcal{B}^*\) depends only on the \(\pmb{\alpha }^*\), and the latter are calculated at every time point following the assumption of maximum entropy (see Barton and de Vladar 2009 for more details on the implementations), we are able to make predictions for the long term response to selection of the expected trait mean and of the expected genetic variance.

\subsection{Approximating the generating function.}

The function $\mathbb{Z}$ (eq. \ref{Eq:DefPartitionFunction}) is a multidimensional integral over the frequencies at the $n$  loci. These integrals are coupled, because the function \(\bar{z}^2\) is not separable into additive terms. Thus, in this section we will focus on an approximation for $\mathbb{Z}$. To begin with, we recall that Wright's formula (1937; eq. 39) implies that we can always express \(\psi (\pmb{p})\) as \(\bar{W}^{2N}\times  \psi _{\text{sf}}\), where \(\psi _{\text{sf}} =\prod _{i=1}^n\exp\left[-4N\!\sigma p_iq_i \right]\left(p_iq_i\right)^{4N\!\mu-1} \) is the distribution of the system free of selection over the trait mean. Thus we include the effects of drift, mutation ($\mu $), and also of selection on the genetic variance ($\sigma $); these are all additive terms. Because \(\psi _{\text{sf}}\) (when normalized) is a product of per-locus distributions, the central limit theorem indicates that with many loci, \(\psi _{\text{sf}}\) will converge to a Gaussian (Appendix 1). Furthermore, the contribution of each locus is typically bimodal, where each peak is close to the borders. Hence, we have \(2^n\) adaptive peaks, which correspond to all the combinations of $\texttt{'}$+$\texttt{'}$/$\texttt{'}$-$\texttt{'}$ alleles. In order to gain accuracy, we expand as a Gaussian not the whole distribution \(\psi _{\text{sf}}\), but rather the distribution around each of the adaptive peaks, by dividing the range of each allele frequency into two regions. Under equal effects, there can be \textit{m} and \textit{n-m} loci near 0 and 1, respectively, weighted by a binomial term (Appendix 1). Each peak will be characterized by its expected mean trait values and variances. Because of the symmetry under equal effects, at each peak the contribution to the expected mean of each allele is equivalent in magnitude, \(M_o\), and of opposite sign for the $\texttt{'}$+$\texttt{'}$ and $\texttt{'}$-$\texttt{'}$ peaks (the trait mean is an odd function of the allele frequencies). Thus, for a given combination (\textit{m,n-m}) the expected trait mean results in \( \langle z\rangle _m\equiv (n-2m)M_o\). Similarly, the variance contributed by each peak is, per locus, \(V_o\), so that for a combination (\textit{m,n-m}) results in \(\text{var}_n\left(\bar{z}\right)\equiv n V_o\) (the variance being an even function of the allele frequencies). In the same way, we may include the expectation of other quantities ($\nu $,\textit{U}, etc., altogether denoted by \pmb{ \textit{a}}) along with their variances and covariances. Defining \(\mathit{c}_m\equiv \left\{\text{cov}_m\left(a_k\right)\right\}_k\), the covariance matrix of \pmb{ \textit{a }}at a given peak with configuration (\textit{m,n-m}), we approximate \(\psi _{\text{sf}}\) as a multivariate Gaussian, namely:

\begin{equation}
\psi _{\text{sf}} = \frac{\mathbb{Z}_o }{(2\pi )^{k/2} }\sum _{m=0}^n \left(
\begin{array}{c}
 n \\
 m
\end{array}
\right)\text{det}\left(\mathit{c}_m\right)^{-1/2}\exp\left[-\frac{1}{2}\left(\pmb{a}-\langle\pmb{a}\rangle _m\right)\cdot \mathit{c}_m^{-1}\cdot \left(\pmb{a}-\langle\pmb{a}\rangle _m\right)^{T}\right] 
\end{equation}

where \textit{k} is the number of variables included in the Gaussian approximation; in practice, we employed \textit{k}=5 variables, namely those required to compute $\mathcal{B}$, which are \pmb{ \textit{a}}=$\{$\(\bar{z},\nu ,m_3,m_4,H\)$\}$. \(\mathbb{Z}_o\) is the normalization constant of this Gaussian approximation, which for equal effects is \(\mathbb{Z}_o\) =\(\left(\tilde{\mathbb{Z}}_o\right)^n\), which in turn is the generating function for $n$  independent single loci at an adaptive peak (Appendix 2):
\begin{equation}
\tilde{\mathbb{Z}}_o = \int _0^{1/2}\exp[4N\!\sigma p  q](p  q)^{4N\!\mu-1}dp .
\end{equation}

After making this approximation, we need to multiply the resulting Gaussian distribution by the effects of selection, as specified by the mean fitness \(\bar{W}^{2N}=\exp\left[2N\!\lambda\bar{z}^2+2N\!\beta \bar{z}\right]\), and integrate on \(\bar{z}\), in order to get $\mathbb{Z}$. Appendix 1 indicates the details and the intermediate steps in the derivation (which we work out for unequal effects, when the complexity of the calculation allows it). In short we find that the generating function is
\begin{equation}
\mathbb{Z} =\frac{\mathbb{Z}_o}{\sqrt{1+4N\!\lambda n V_o}}\exp\left[\frac{2 N\!\beta^2}{1+4N\!\lambda n V_o}\right] \times \sum _{m=0}^n \left(
\begin{array}{c}
 n \\
 m
\end{array}
\right)\exp\left[-\frac{4 m n}{1+4N\!\lambda n V_o}\left(N\!\beta -2 N\!\lambda (n-2m)M_o\right)\right]
\end{equation}

The expectations of the macroscopics follow from the derivatives of Log[$\mathbb{Z}$] w.r.t. the parameters \pmb{ $\alpha $} as explained above, all of which can be written explicitly (see Appendix 1). In practice because $n$ is large, the binomial terms
\(
\left(
\begin{array}{c}
 n \\
 m
\end{array}
\right)
\) are computed by approximating them by a Gaussian, and the binomial sum by an integral, thus:
\begin{equation}
\sum _{m=0}^n  \left(
\begin{array}{c}
 n \\
 m
\end{array}
\right) F_m \simeq \frac{2^n}{\sqrt{n \pi }}\int _{-\infty }^{\infty }\exp\left[-2\left.(m-n/2)^2\right/n\right] F_{(m)} dm
\end{equation}
which allows computations that can be handled numerically (Appendix 1). In the following sections, we study the accuracy of our approximations.

\section{RESULTS}

\subsection{Numerical experiments}

Exact analysis of the distribution of allele frequencies is in theory possible using a diffusion equation approach (Crow and Kimura 1970, ch. 7). But for many coupled loci, as in this case of SS, the time-dependent analytic solution is unknown, and its numerical calculation is not feasible --short of Monte Carlo methods that would amount to simulation of the original problem. Thus, the possibilities to check our method are reduced to either simulating allele frequencies, or to employ individual based simulations. We take the first approach, that is to draw multiple realizations of eq. (\ref{Eq:Wright-FisherModel}), each giving a stochastic path of the allele frequencies, employing in our purposes an Euler scheme with time step $\Delta $t=0.01, and drawing deviates from a Gaussian with variance \(\text{$\Delta $t} p  q/2N\) for the stochastic term modeling genetic drift (e.g. Higham 2001). From these paths, we compute the trait mean and its genetic variance. Then, we can average over multiple realizations in order to estimate the expectations of these variables. We ignore linkage disequilibrium, but know that this is negligible for recombination rates \textit{r }$>\!>$ $\mathit{s}$ (B{\" u}rger 2000, ch. VI.7). Thus we work entirely from the diffusion approximation to the evolution of allele frequencies, assuming that selection is weak ($\mathit{s}$ $<\!<$ \textit{r},1). Our statistical mechanical approach assumes that evolution starts from an initial equilibrium; it does not allow arbitrary initial distributions that would give arbitrary outcomes. Therefore, for a fair comparison, we need to start the numerical realizations also from an equilibrium, drawing an initial set of allele frequencies from the theoretical marginal distribution. Because the joint distribution of allele frequencies and the product of the marginals are not the same, we still allow the process to relax to the stationary state for a few thousands of generations (typically, \(\sim 10^5\)), until numerically we obtain that \(\bar{z}\simeq  z_{\text{opt}}\). Only then we allow selection to change. From this moment, when evolution proceeds, we average and record the values to compare them later with the SM estimate.\\
In order to perform this comparison, we perform a significance test. The null hypothesis \(\mathcal{H}_0\) is that the numerical calculations are not significantly different from the SM expectations. Thus to support our theory, we would like to accept \(\mathcal{H}_0\). Because the numerical expectations are averages over many realizations, these are normally distributed. From the SM theory we know the expectation and the variance of the macroscopics that we want to test, namely, the trait mean and the genetic variance. Summing-up, we know the expected values and the variance and thus, ideally the residuals should be normally distributed. Hence, a \textit{z}-test for the goodness of fit is enough for our purposes (Cox and Hinkley 1974, ch. 3). The \textit{z-}statistic is \(z=\sum _{\tau =1}^T \left(\bar{A}_{\tau }-\left\langle A^*\right\rangle \right)/ \text{var}\left(A^*\right)\), where \(\bar{A}\) are the averages over the simulations, \(\left\langle A^*\right\rangle \) is a short-hand notation for \(\langle A\rangle _{\left(\alpha ^*(t)\right)}\) (the expectation of \textit{A} evaluated at the local forces at time $\tau $), \textit{T} is the total number of time points, and \textit{df=T}-1 are the degrees of freedom. We arbitrarily set the significance level to 99$\%$. In addition, we also compare the maximum observed deviation, \(\text{Max}_{\tau }\left(\bar{A}_{\tau }-\left\langle A^*\right\rangle \right)\) with the standard deviation from the SM at that same time point. Respectively, these two statistics quantify for the SM approximation the \textit{accuracy} (that is, how close are the predictions to the true value) , and the \textit{precision} (that is, how much the observations deviate from its expected value). In turn these two measures give us a grip to judge whether our approximation deviates significantly from the $\texttt{'}$true$\texttt{'}$ values (from the simulations), and also whether significant deviations are biologically meaningful. As we will see in some examples, some systematic and statistically significant deviations may occur.

\subsection{Shifting optimum}

\begin{figure}[t]
\begin{center}
\includegraphics[scale=0.56]{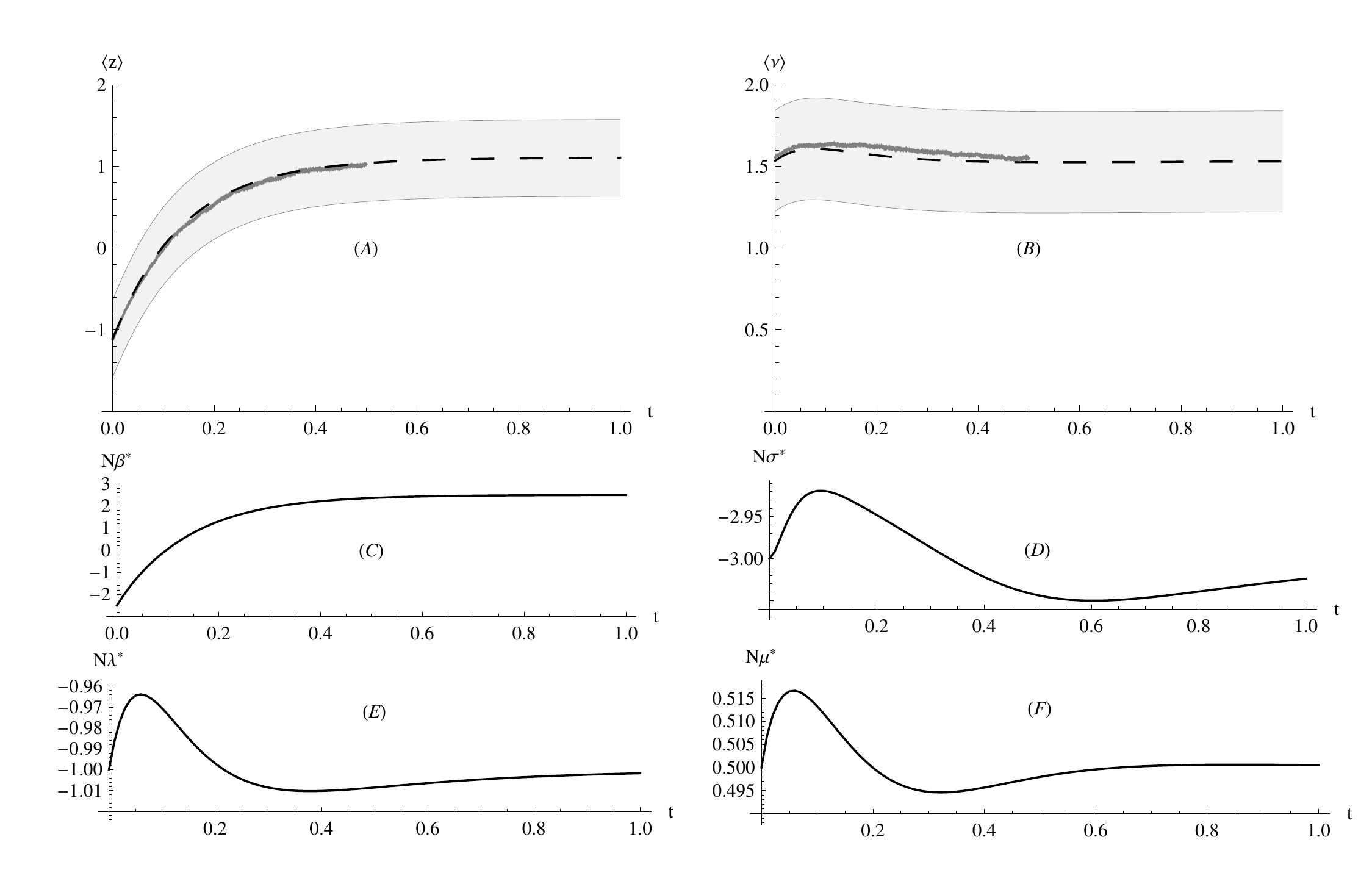}
\caption{Evolutionary dynamics when the optimum changes abruptly, shifted at time t=0 from -0.75 to 0.75 (which corresponds to $N\beta $ =-2.5 to$N\beta $=2.5, see panel C). The trait consists of  $n$=5 loci of equal effect. Expectations (black dashed
lines) of (A) a polygenic trait and (B) its genetic variance. The gray regions cover $\pm $ the standard error (root mean squared deviation of the
variance of the macroscopics). The change in the genetic variance is at most of 4.85$\%$ of the initial value, while its standard error is 20.5$\%$.
The gray solid lines are averages of the numerical realizations (with population size of $N=100$; 1000 replicas were employed). The goodness
of fit (one tail chi-square with 51 degrees of freedom) accepts the null hypothesis (the simulation points are random samples of the SM distributions)
in both cases: (A) \(z=-0.058,p=0.68\); the maximum deviation is 12.19$\%$ of the standard error (0.47) ; (B) \(z=0.12,p=0.40\); the maximum deviation
is 18.27$\%$ of the standard error (0.31) . (C-F) Evolution of the local forces. Notice the short range of change on (D-F), these forces remain practically
constant. $N\!\lambda $=-1.0, $N\!\sigma $=-15 and $N\!\mu $=0.3.}
\label{Fig:OptimumAbruptChange5loci}
\end{center}
\end{figure}

\begin{figure}[t]
\begin{center}
\includegraphics[scale=0.47]{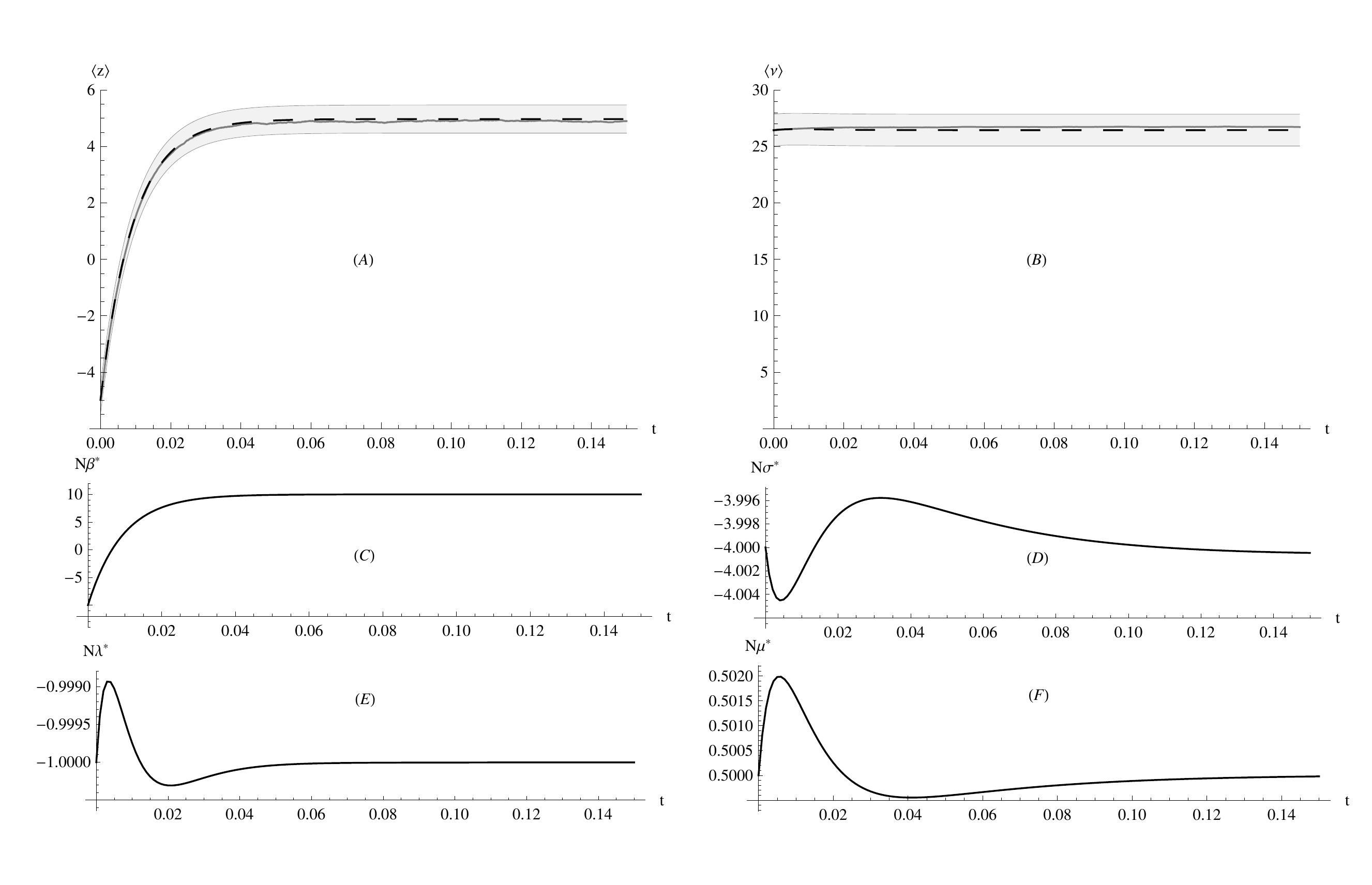}
\caption{Evolutionary dynamics when the optimum changes abruptly, shifted at time t=0 from -5 to 5 (which corresponds to $N\!\beta $ =$\pm $10, that is about 5$\%$ of the total range; see panel C). The trait consists of  $n$=100 loci of equal effect. Expectations
(black dashed lines) of (A) a polygenic trait and (B) its genetic variance. The gray regions cover $\pm $ the standard error (root mean squared deviation
of the variance of the macroscopics). The change in the genetic variance is at most of 0.4$\%$ of the initial value, while its standard error is
10.3$\%$. The gray solid lines are averages of the numerical realizations (with $N=100$; 500 replicas were employed). The goodness of fit
(151 degrees of freedom) accepts the null hypothesis in both cases: (A) \(z=-1.14,p=0.086\); the maximum deviation is 26.5$\%$ of the standard error
(0.50) ; (B) \(z=0.18,p=0.03\); the maximum deviation is 22.19$\%$ of the standard error (1.41) . (C-F) Evolution of the local forces. Notice the
short range of change on (D-F), these forces remain practically constant. $N\!\lambda $=-1.0, $N\!\sigma $=-4 and $N\!\mu
$=0.5. Otherwise as in fig. \ref{Fig:OptimumAbruptChange5loci}.}
\label{Fig:OptimumAbruptChange100loci}
\end{center}
\end{figure}

The most radical test of our approximation is when selection changes abruptly. For example, a sudden shift of the optimum would trigger a quick response of the trait, and a major reconfiguration of the genetic states. The prediction of the change of the trait mean and of the genetic variance is thus not a trivial task. In turn, our approximation allows us to estimate the change of their expectations, which gives a rather robust prediction of their evolutionary course. Figures \ref{Fig:OptimumAbruptChange5loci} and \ref{Fig:OptimumAbruptChange100loci} present two examples of this situation. We compare this response with intensive simulations of the Wright-Fisher model (eq. \ref{Eq:Wright-FisherModel}), for two distinct numbers of loci: five and a hundred. The response of the mean is quicker for traits with more loci. In this case of equal effects, the reason is clear: the expectations of the trait mean and the genetic variance are $\texttt{'}$extensive$\texttt{'}$, meaning that their values are proportional to the number of loci, $n$ . The expectations are not strictly linear with $n$ , but it is nevertheless a positive dependence. Thus, traits composed of a larger number of loci have larger genetic variance, and therefore respond faster to selection. This is expected from the formula from the Stochastic House of Cards model (SHoC; B{\" u}rger 2000, pp. 270, eq. 2.8) that predicts that the expected genetic variance will be
\begin{equation}
\label{Eq:GeneticVarianceHoC}
\langle \nu \rangle _{\text{SHoC}}=\frac{2n \gamma ^2N  \mu }{1+\gamma ^2|N  \sigma |}
\end{equation}
(in this case, we set the effect to $\gamma $=1). Thus increasing the number of loci, increases the genetic variance. The precision of our approximation does not seem to depend critically on $n$ , except for a very low number of loci ($n$  between 3 and 5, depending on the parameters, results not shown), where the Gaussian approximation fails. Yet, the predictions of the macroscopics are in very good agreement with the numerical expectations, even for $n$  as low as 5.

\begin{figure}[t]
\begin{center}
\includegraphics[scale=0.5]{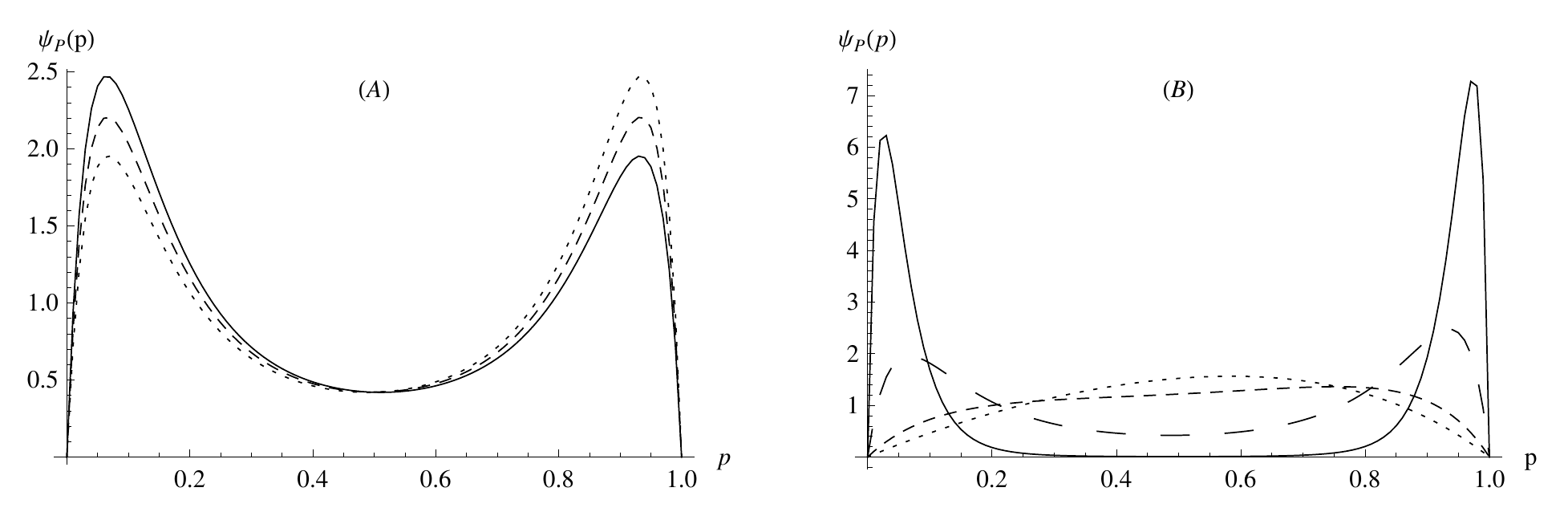}
\caption{Marginal distribution of frequencies of an allele. (A) Changing the intensity of selection over the trait modulates
the height of the adaptive peaks (other things being equal, N$\sigma $=-3.0) resulting in pronounced changes in the expectation of the trait mean
but with weak changes (even a relatively constant value) of the expectation of the genetic variance. Solid line: $N\beta $=2.5, \(\left\langle \bar{z}\right\rangle
=1.21, \langle \nu \rangle =1.12\). Dashed line: $N\beta $=0, \(\left\langle \bar{z}\right\rangle =0, \langle \nu \rangle =1.13\). Dotted line: $N\beta
$=-2.5, \(\left\langle \bar{z}\right\rangle =-1.21, \langle \nu \rangle =1.12\). (B) Changing the intensity of selection over the genetic variance
modulates the position of the adaptive peaks (other things being equal, $N\beta $=2.5) resulting in pronounced changes in the expectation of the
genetic variance, but with weak changes on the expectation of the trait mean. Dotted line: N$\sigma $=0, \(\left\langle \bar{z}\right\rangle =1.18,
\langle \nu \rangle =8.04\). Short-dashed line : N$\sigma $=-1.0, \(\left\langle \bar{z}\right\rangle =1.19, \langle \nu \rangle =7.52\). Large-dashed
line: N$\sigma $=-4.0, \(\left\langle \bar{z}\right\rangle =1.22, \langle \nu \rangle =5.32\). Solid line: : N$\sigma $= -10.0, \(\left\langle \bar{z}\right\rangle
=1.36, \langle \nu \rangle =2.20\). In all cases N$\mu $=0.5, N$\lambda $=-1.0, and the trait is composed of 20 loci of equal effects.}
\label{Fig:MarginalDistributions}
\end{center}
\end{figure}

In these examples, we find that the expected genetic variance remains practically unchanged, even though $\nu $ fluctuates wildly in any one realization. In fact, in the two cases in figs. \ref{Fig:OptimumAbruptChange5loci} and \ref{Fig:OptimumAbruptChange100loci} , \(\langle \nu \rangle \) is variable, but achieves at most an increase of 1$\%$. This makes it not only hard but to some extent pointless to aim for an accurate numerical or --more critically-- empirical estimation of \(\langle \nu \rangle \) (Fowler and Whitlock 2002), in part because we would need an unrealistic number of observations. The variance of a set of independent (numerical or empirical) measurements of $\nu $ is of comparable magnitude to the mean range of response of such measurements, so any change in $\nu $ will be obfuscated by the effects of drift.\\
However, the nearly constant patterns of the expectations of $\nu $ should not be confused with a constancy of $\nu $ itself. The latter will of course fluctuate around its expectation because of genetic drift. Accordingly, we evaluated the deviations from the expectations by computing Var($\nu $) (which follows from the derivative \(\left.\partial ^2\log[\mathbb{Z}]\right/\partial 4N\!\sigma^2\), an element of the matrix \(\mathcal{C}\)), and from it, the root mean square deviation (RMSD). The RMSD shows the range in which the individual paths are most likely to occur. These bounds are also shown in the figures 3 and 4, exposing the fact that the effects of drift are big compared to the change in the expectations, but still in a narrow percentile range of the actual standing variation. This indicates that in practice $\nu $ can be considered constant. Such small changes in the genetic variance and its expectation justify the classical approaches like the breeder's equation, which relies on constant heritabilities for long term predictions (Lande 1979). For these kind of situations, a complicated mathematical machinery like the one introduced here seems unjustified. Nevertheless, we can still apply our calculations in order to compute other aspects that don't arise naturally from the classical theory. For example, we can easily estimate the expected variance of the set of measurements alluded above (Fowler and Whitlock 2002). This not only gives a rule of thumb of when to expect genetic variance to remain constant, but also provides a way to falsify the SM approach. With increasing numbers of loci (a) the range of change of $\nu $ becomes smaller (there is larger mutational load), and (b) the range of fluctuation (i.e. the standard error) increases. Thus, we are even less able to detect changes in the expectations.\\
These calculations give some insight into why \(\langle \nu \rangle \) hardly varies in response to selection. As figs. \ref{Fig:OptimumAbruptChange5loci} and \ref{Fig:OptimumAbruptChange100loci}  suggest, drift would be the main driving force for the changes in genetic variance. Why isn't the change in \(\langle \nu \rangle \) more pronounced?\\
Thinking in terms of the distribution of allele frequencies (fig. \ref{Fig:MarginalDistributions} , showing the marginal distribution at one locus), we find that the distribution is bimodal. The positions of the adaptive peaks determine the genetic variance, which depends only weakly on their height. The relative height of these peaks, on the other hand, determine the value of the trait mean, but depend weakly on their position. Thus, because these two quantities (\(\bar{z}\) and $\nu $) are practically uncoupled, it is possible to select over them more or less independently. Selection on the trait mean will cause multiple jumps among the peaks (Barton 1986). But the position of the peaks remain unchanged, because they are equidistant to the saddle, which is situated at an intermediate frequency \(p_s\)=1/2 (this is because there is no dominance). Consequently, because the genetic variance is symmetric w.r.t the allele frequencies, few $\texttt{'}$-$\texttt{'}$ alleles with frequency \textit{p} (and thus $\texttt{'}$+$\texttt{'}$ alleles with frequency. 1-\textit{p}) result in as much variance as numerous $\texttt{'}$-$\texttt{'}$ alleles with frequency 1-\textit{p} (and $\texttt{'}$+$\texttt{'}$ alleles of frequency \textit{p}). So, although the alleles are jumping from one peak to the other, their frequencies change always from \textit{p} to 1-\textit{p} (or vice-versa), affecting the proportions of $\texttt{'}$+$\texttt{'}$/$\texttt{'}$-$\texttt{'}$ alleles at each peak, and thus the trait, but without affecting the genetic variance. However, the intermediate frequencies are transiently populated by the jumping alleles, so a slight increase in the genetic variance should be observed. This change, as a matter of fact is predicted by the SM but it is generally only a few percent of the standard error, and often too small to be detected. Notice however, that the force related to the genetic variance (that is, $N\!\sigma $) shows a small change. But the scale is so narrow that it can practically be considered constant. This is of course supported by the SHoC expectation of the genetic variance, eq. (\ref{Eq:GeneticVarianceHoC}), where the expectation of the genetic variance does not depend on \(\left\langle \bar{z}\right\rangle \).

\subsection{Effect of the number of loci on the rates of evolution.}

As suggested by comparing figs. \ref{Fig:OptimumAbruptChange5loci} and \ref{Fig:OptimumAbruptChange100loci}, with 5 and 100 loci, respectively, the amount of standing genetic variation is proportional to the number of loci $n$ . This insinuates that traits composed of more loci should adapt quicker to ecological changes. But, is there any relationship between these rates of adaptation and the number of loci? As a matter of fact, if we measure time relative to the expected number of mutations (that is 2$n$ $\mu $\textit{t}) the time to reach an equilibrium are comparable for different number of loci. In this time scale, we find that the genetic variance is slowed down, but it also reaches higher levels for traits composed of larger numbers of loci (fig. \ref{Fig:DeterministcExample} in the electronic supplementary material, ESM 1).  In the limit of infinite loci, we would have (1) constant genetic variance, but (2) its value would diverge, leading to {`}instant{'} adaptation. A solution to this paradoxical limit, comes from rescaling the effects of the alleles with the number of loci (Bulmer 1980, Turelli and Barton 1990). Thus in making \(\gamma \sim 1\left/\sqrt{n}\right.\), the genetic variance is regularized: as the number of loci increases, its rate of change go to zero, and it converges to a constant value, that is, the infinitesimal model (Bulmer 1980). These circumstances are depicted in fig. 2 of the ESM 1.

\subsection{Moving optimum}

\begin{figure}[t]
\begin{center}
\includegraphics[scale=0.47]{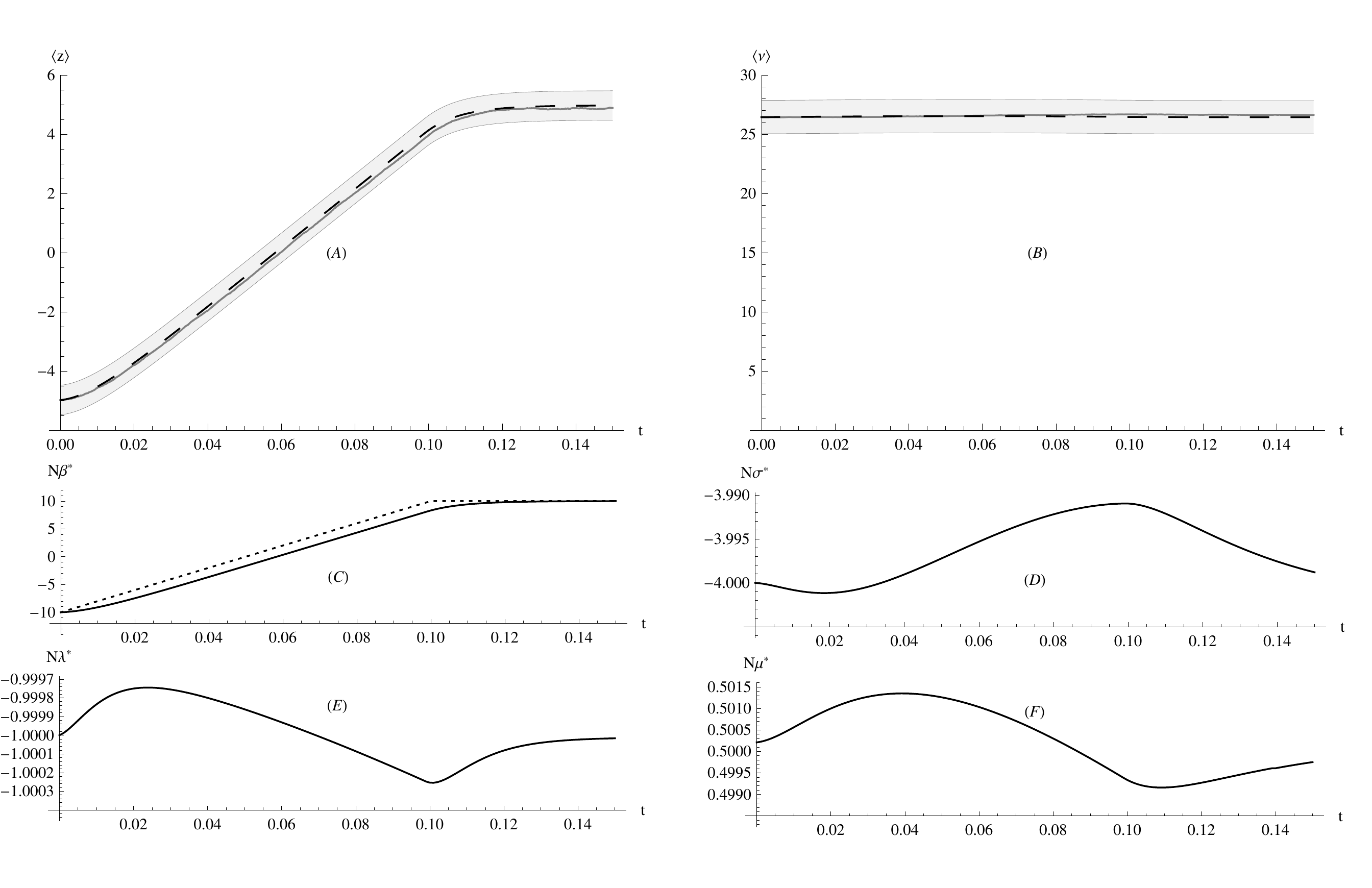}
\caption{Evolutionary dynamics when the optimum moves gradually from -5 to 5 (which corresponds to $N\!\beta $ =$\pm
$10, see panel C, dotted line) at a rate of \(5 10^{-3}\) units per generation. The trait consists of  $n$=100 loci of equal effect. Expectations
(black dashed lines) of (A) a polygenic trait and (B) its genetic variance. The gray regions cover $\pm $ the standard error (root mean squared deviation
of the variance of the macroscopics). The gray solid lines are averages of the numerical realizations (with$N=100$; 500 replicas were employed).
The goodness of fit (151 degrees of freedom) rejects the null hypothesis for the trait mean, and accepts it for the genetic variance: (A) \(z=-0.21,p=0.009\);
however, the maximum deviation is 37.31$\%$ of the standard error (0.50); (B) \(z=0.067,p=0.41\); the maximum deviation is15.94$\%$ of the standard
error (1.41). (C-F) Evolution of the local forces. Notice the short range of change on (D-F), these forces remain practically constant. $N\!\lambda $ = -1, $N\!\sigma $ = -4, $N\!\mu $ = 0.5. Otherwise as in fig. \ref{Fig:OptimumAbruptChange5loci}.}
\label{Fig:MovingOptimum100lociShortRange}
\end{center}
\end{figure}

\begin{figure}[t]
\begin{center}
\includegraphics[scale=0.46]{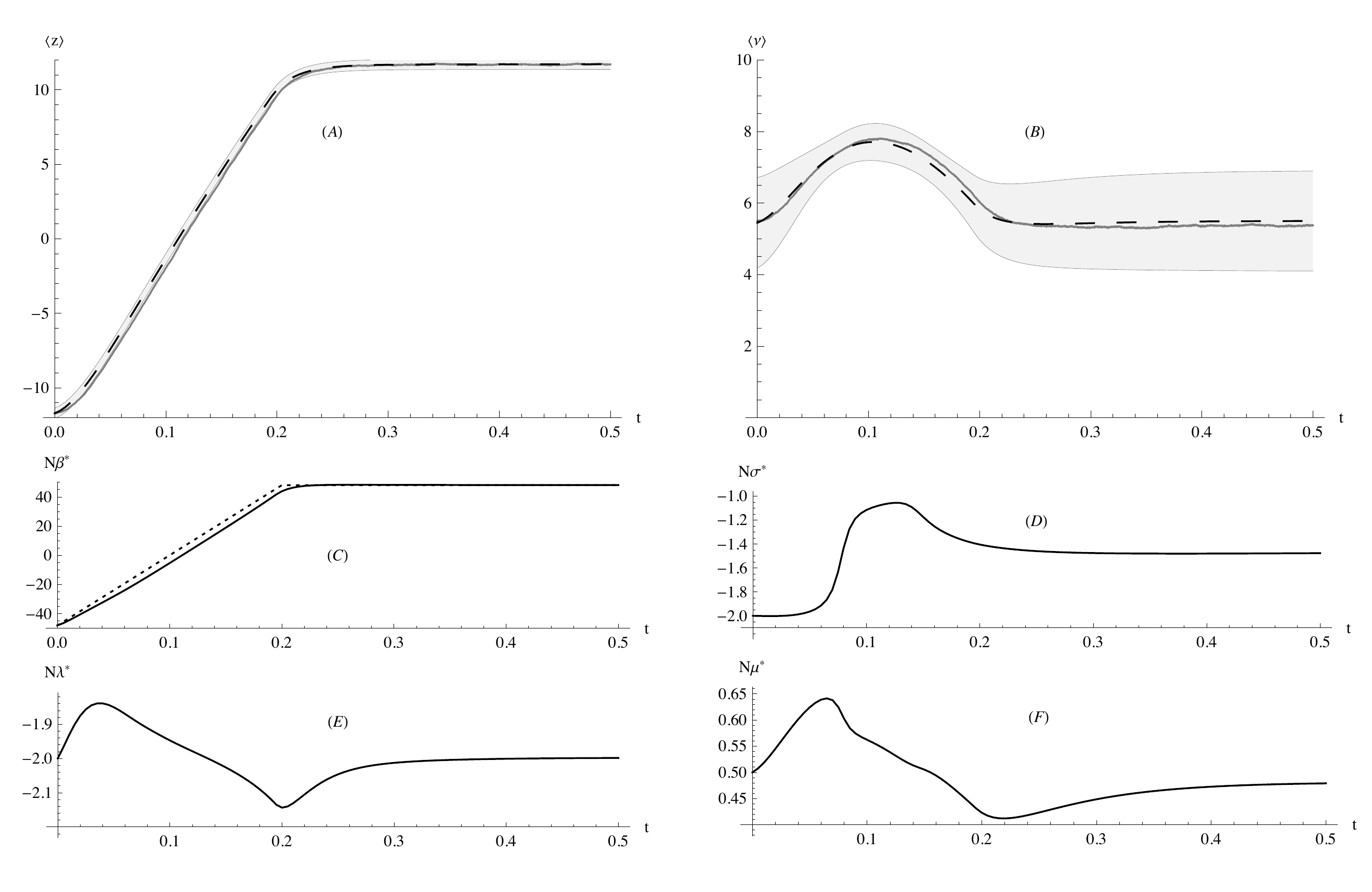}
\caption{Evolutionary dynamics when the optimum moves gradually from -12 to 12 (which corresponds to $N\!\beta $
=$\pm $48, 60$\%$ of the total range; see panel C, dotted line) at a rate of \(5 10^{-3}\) units per generation. The trait consists of  $n$=20
loci of equal effect. Expectations (black dashed lines) of (A) a polygenic trait and (B) its genetic variance. The gray regions cover $\pm $ the
standard error (root mean squared deviation of the variance of the macroscopics). The gray solid lines are averages of the numerical realizations
(with $N=100$; 500 replicas were employed). The goodness of fit (101 degrees of freedom) rejects the null hypothesis for the trait mean,
and accepts it for the genetic variance: (A) \(z=-0.64,p<10^{-9}\); the maximum deviation is 171$\%$ of the standard error (0.36); (B) \(z=0.06,p=0.54\);
the maximum deviation is 41$\%$ of the standard error (0.69). (C-F) Evolution of the local forces. Notice the short range of change on (D-F), these
forces remain practically constant. $N\!\lambda $ = -2, $N\!\sigma $ = -2, $N\!\mu $ = 0.5. Otherwise as in fig.
 \ref{Fig:OptimumAbruptChange5loci}.
}
\label{Fig:MovingOptimum20loci}
\end{center}
\end{figure}

Another example that was alluded in the introduction and which is of general interest is when the optimum is steadily shifting its position (B{\" u}rger 2005; Waxman 2005; Kopp and Hermisson 2007; Sato and Waxman 2008; Kopp and Hermisson, 2009). Figure \ref{Fig:MovingOptimum100lociShortRange}  presents a comparable situation to that of fig. \ref{Fig:OptimumAbruptChange100loci}, but where the optimum is moving slowly. Again, we do not detect major changes in \(\langle \nu \rangle \). Here, the trait is being selected in a range that is much less than its total range of response ($\pm $\textit{n, n}=100 in this case, while the optimum shifts in a comparably short range, between $\pm $5). However, if we consider a larger range, comparable to $n$  (depicted in fig. \ref{Fig:MovingOptimum20loci} ), then at the extremes we force fixation of some alleles, and the variance changes significantly. Notice that during evolution, the range of change of the effective selective values (what we termed forces) can change notably (fig. \ref{Fig:MovingOptimum20loci} ). But if we compare to other scenarios (e.g. fig. \ref{Fig:MovingOptimum100lociShortRange} ), the forces can be virtually constant.\\
Despite the previous examples, the polygenic dynamics when the optimum is steadily moving can be very complicated, because as the optimum shifts the allelic combinations that match the optimum continuously change with dramatic microscopic modifications (Kopp and Hermisson 2007 2009). The result is that the path of $\nu $ is erratic in the deterministic case, because the alleles at different loci keep sweeping and their frequency changing, inflating and deflating the genetic variance in an almost chaotic fashion. This is all smoothed out when drift is present and we average over an ensemble. Thus, a more serious test for our theory would be to consider situations where there is considerable chance that the dynamics get stuck in local minima.

\subsection{Selection on the genetic variance.}

\begin{figure}[t]
\begin{center}
\includegraphics[scale=0.4]{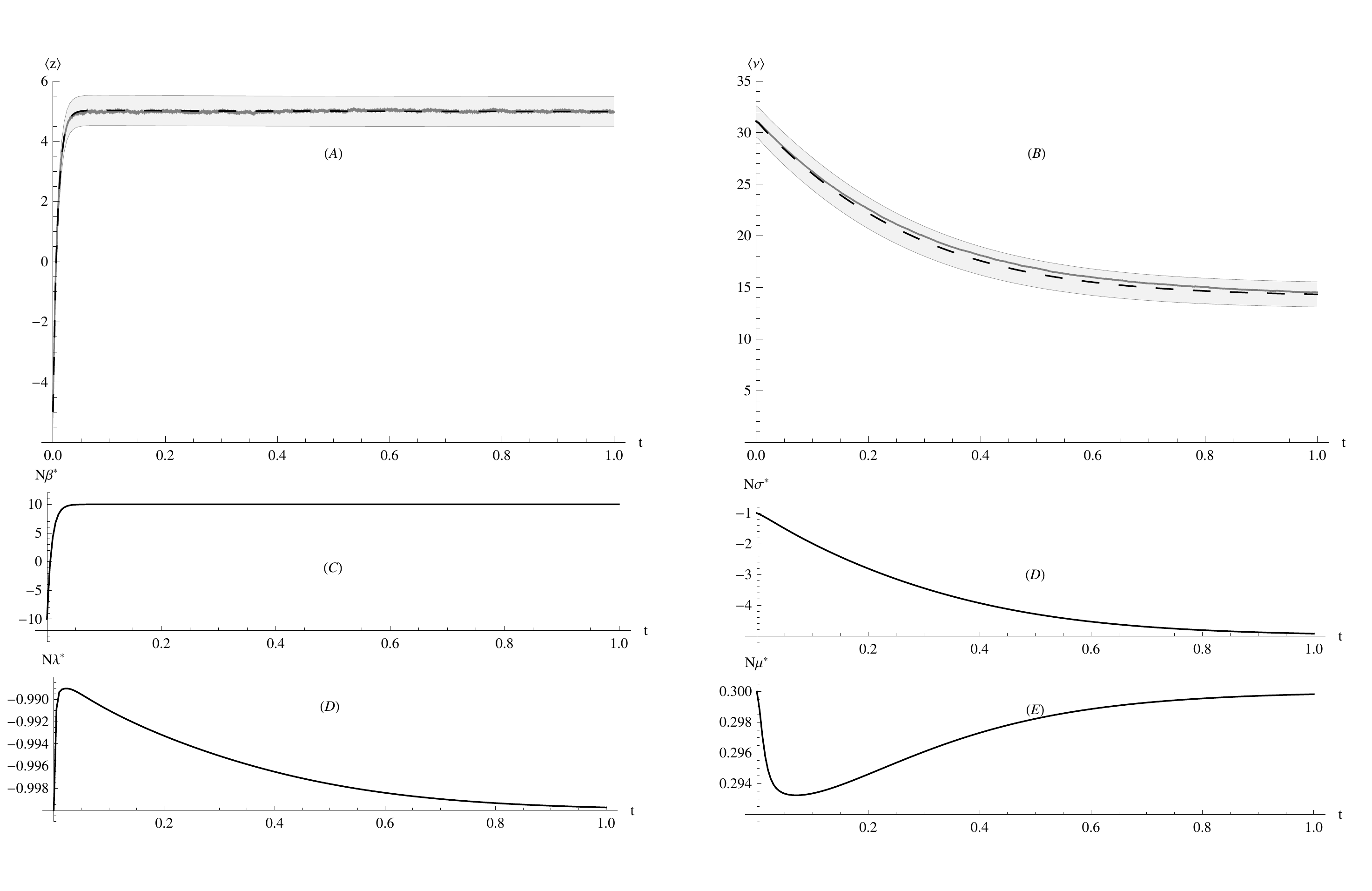}
\caption{Evolutionary dynamics when the optimum changes abruptly, shifted at time t=0 from -5 to 5 (which corresponds to $N\!\beta $ =$\pm $10, see panel C), and simultaneously, selection against the genetic variance is increased from $N\!\sigma $=-1 to $N\!\sigma $=-5. The trait consists of  $n$=100 loci of equal effect. Expectations (black dashed lines) of (A) a polygenic trait and (B) its
genetic variance. The gray regions cover $\pm $ the standard error (root mean squared deviation of the variance of the macroscopics). The change
in the genetic variance is at most of 0.4$\%$ of the initial value, while its standard error is 10.3$\%$. The gray solid lines are averages of the
numerical realizations (with $N=100$; 500 replicas were employed). The goodness of fit (151 degrees of freedom) accepts the null hypothesis
for the trait mean, but rejects it for the genetic variance: (A) \(z=-0.011,p=0.87\); the maximum deviation is 17.19$\%$ of the standard error (0.50);
(B) \(z=0.28,p<10^{-4}\); the maximum deviation is 39.77$\%$ of the standard error (1.37). (C-F) Evolution of the local forces. Notice the short
range of change on (D-F), these forces remain practically constant. $N\!\lambda $ =-1, $N\!\mu $ = 0.3. Otherwise as in fig. \ref{Fig:OptimumAbruptChange5loci}.}
\label{Fig:SelectionGeneticVariance}
\end{center}
\end{figure}

In part, whether our predictions make biological sense, is a matter of time scale. In the first examples, genetic drift eliminated genetic variance at a similar rate to its production by mutation. Thus, the rate of change of \(\langle \nu \rangle \) was very small. We can obtain a different picture if the genetic variance is forced to change along with, or instead of, selecting over the trait. Under stabilizing selection, the mean fitness is determined by two variance measures: the genetic variance, and the squared deviation of the trait mean w.r.t. the optimum. Both of them are under selection with strengths $N\!\sigma $ and $N\!\lambda $, respectively. The canonical way to represent stabilizing selection (Lynch and Lande 1993) will assume that \(N  \sigma =N  \lambda =\mathit{s}/2\). We relax this constraint, and allow selection to act independently over $\nu $ and over the deviation from the optimum. At the moment, we focus on the first alternative.\\
The scenario that we study, is that when a population is initially under selection-mutation-drift equilibrium where selection acts over a quantitative trait, but with little strength on the variance. Suddenly, selection against the variance increases, and the optimum changes sign. This implies a radical reconfiguration of the adaptive landscape. In this case, will the SM approximation be accurate? An example of this scenario is demonstrated in fig. \ref{Fig:SelectionGeneticVariance} . We find that there is a good agreement with the simulations.\\
As explained above, selection over the trait mean and over the genetic variance are practically uncoupled. Thus if the optimum moves without affecting the strength of selection, the trait mean can freely evolve without significant changes on the genetic variance (figs. \ref{Fig:OptimumAbruptChange5loci},  \ref{Fig:OptimumAbruptChange100loci}, \ref{Fig:MovingOptimum100lociShortRange}). However, we have now studied a more general situation when both forces, the optimum and the selection strength, change. Figure \ref{Fig:SelectionGeneticVariance} shows that the pattern of change of the force \(N\!\beta^*\) is similar to the one in fig. \ref{Fig:OptimumAbruptChange100loci} where a similar situation was studied, except that the genetic variance did not change. This suggest that even in this case (fig. \ref{Fig:SelectionGeneticVariance} ) where both selection and the optimum shift, the evolution of the genetic variance and of the trait mean are uncoupled. This is unexpected because, unlike the previous example, the adaptive peaks are being relocated. This rearrangement of the microscopic space, which originally was nearly peaked close to 1/2 but which suddenly became bi-modal, partitioned the occurring alleles closer to the borders of fixation --these configurations that diminish the genetic variance. Simultaneously, a high frequency of $\texttt{'}$+$\texttt{'}$ alleles was produced, and at another locus a high frequency of $\texttt{'}$-$\texttt{'}$ alleles, in such proportions that maintain low genetic variance, while allowing the trait mean evolve to increase to its new optimum value. \\
Under the quasi-equilibrium approximation, the SHoC formula (eq. \ref{Eq:GeneticVarianceHoC}) would hold for any time if we would substitute $N\!\mu $ and $N\!\sigma $ for the effective forces \(N\!\mu^*\) and \(N\!\sigma^*\). Thus the expectation of the genetic variance would only change if any of these change. If the optimum value is displaced without significantly changing \(N\!\mu^*\) or \(N\!\sigma^*\), there would be no reason to expect a change in \(\langle \nu \rangle _{\text{SHoC}}\). A marginal issue here, is whether the predictions from eq. (\ref{Eq:GeneticVarianceHoC}) are in agreement with the SM expectations. For example, for the initial state depicted in fig. \ref{Fig:SelectionGeneticVariance} , the numerical averages (from the Wright-Fisher simulations) for the genetic variance give \(\hat{\nu }_{\text{WF}}=\)31.13 $\pm $ 1.65, the SHoC gives \(\langle \nu \rangle _{\text{SHoC}}\) = 30.00, whilst the SM gives \(\langle \nu \rangle _{\text{SM}}\) = 31.08 $\pm $1.5. For the end equilibrium, we have \(\hat{\nu }_{\text{WF}}=\)14.08$\pm $1.36 , and the models give \(\langle \nu \rangle _{\text{SHoC}}\) = 10.00, whilst \(\langle \nu \rangle _{\text{SM}}\) = 14.12$\pm $1.21. There seems to be a decent agreement between both predictions. However, the agreement between the SHoC and the SM expectations does diverge as $N\!\sigma \to $0 (data not shown).

\subsection{Further dynamical modes}

Quantitative traits, and the genetic variance, under stabilizing selection can also evolve by other means than selecting over the variance, and changing the position of the optimum. The statistical mechanics allows \(\bar{z}\) and $\nu $ to vary by changes in $N\!\lambda $. Recalling that this parameter penalizes the deviations of the trait mean from the optimum, varying it should not affect the value of \(\bar{z}\). In fact, if we keep the optimum at a fixed place, and say, we reduce $N\!\lambda $, the \(\left\langle \bar{z}\right\rangle \) and $\langle \nu \rangle $ remain almost unchanged. Instead, there is an increase in the standard deviation of \(\bar{z}\). Thus this parameter $N\!\lambda $ determines how much \(\bar{z}\) deviates from its the expected values. Surprisingly, the marginal distribution of allele frequencies remains practically unchanged. Despite that an adaptive peak is not the same as the marginal distribution of an allele, the mean and variance at each adaptive peak do not depend on $N\!\lambda $ (see appendix 2), thus altering its value will not affect the positioning and relative heights of the adaptive peaks. However, $N\!\lambda $ would make the distribution more or less concentrated at such peaks. Thus although the expectations remain the same, populations with larger $N\!\lambda $ would show population means and variances more scattered than other population (or states) where $N\!\lambda $  is smaller (for numerical examples, see figs. 4 and 5 of the ESM 1).\\
Another way in which populations can evolve, is by changing the mutation rate. An increase (decrease) in the number of mutants will inflate (deflate) de expected genetic variance, but will not affect the trait mean{'}s expectation. However, if the mutation rate approaches the $\mu =1/4N@$, the SM method is inaccurate (this will be discussed in more detail in \S 4.7) .

At such point, the borders of the distribution (i.e. \textit{p}=0,1) become absorbing, leading to a failure of the local equilibrium dynamics (figs. 6 and 7 of the ESM 1). Close to the boundaries, two effects are present: (1) drift is much more powerful than selection in fixating alleles, and (2), mutation does not produce enough alleles to keep the distribution away from fixation. Towards the center of the distribution, selection is more effective than drift in maintaining polymorphisms. The rate at which each process occurs have different time scales: fixation towards at the borders happen very fast, whereas diffusion away from the borders is much slower. The local equilibrium dynamics cannot cope with both time scales at the same time, and the SM approximation fails (Barton and de Vladar 2009).

\subsection{Some negative results: failure of the SM approximation.}

\noindent \pmb{ Failure at low mutation rates}. In our previous paper, where we dealt with directional selection (Barton and de Vladar 2009), the SM framework had to be modified for low mutation rates ($N\!\mu <$1/4). At these low mutation rates, there is a major qualitative change in the distribution of allele frequencies, since the fixation states suddenly become absorbing. This is also true in the present case of stabilizing selection. Thus, when mutation rates are close to this critical point$N\!\mu $=1/4, the local equilibrium underestimates the rates of change (e.g. the genetic variance), and the accuracy of our predictions is lost. In the ESM 1 (Sect. 5), we show two examples, with the optimum abruptly increased and for a moving optimum, where the predictions deviate significantly.\\
When $N\!\mu <$1/4 the macroscopics that characterize the system change, in what is called a phase transition. In this case, the genetic heterozygosity \textit{U}, has to be dropped, and the dynamics of the other macroscopics must be computed from the jump processes (using a master equation, instead of diffusion; Barton and de Vladar 2009). This works well for directional selection, but would lead to a different set of equations for the local variables and for the macroscopics compared with the high mutational input case. Nevertheless, the philosophy is the same: to maximize entropy under the constraint of the appropriate macroscopics, and approximate the dynamics using the local equilibrium Ansatz.

\noindent \pmb{ Failure when selection on the genetic variance is strong.} We also found that when selection on the genetic variance is very strong, 4$|$$N\!\sigma |>\!>$1, we also lose accuracy. In the ESM 1, we illustrate this situation for abrupt changes and gradual moves in the optimum trait value. The reason for the failure seems to be that when the valleys between the adaptive peaks are deep, the microscopic distribution approaches a stationary distribution very slowly, so that our central assumption that $\psi $ is nearly stationary, fails. Again, a jump process seems to be a more accurate model than diffusion, which we are eager to apply for stabilizing selection. We will come back to these explanations later.

\subsection{Hysteresis}

The quasi-equilibrium assumption will hold when the changes in the microscopic states are faster than the macroscopic changes. Because of the complexity of the adaptive landscape, a given macroscopic state can stand at different adaptive peaks. Although these peaks may be indistinguishable from the macroscopic point of view, they will have an effect on the rates of change of the latter. Because the transitions among different adaptive peaks may in general happen at different rates, then the final macroscopic state -- after a peak shift -- will depend on the particular microscopic configuration at the time of the jump. This is remarkable, because the stationary distribution is unique, and is determined solely by \(\langle \pmb{A}\rangle \). An immediate implication is that the trajectory that the macroscopics take during evolution, depends on their previous history. That is, the dynamics show hysteresis. Thus, even when the microscopic variables are averaged out, there can be memory in the macroscopic dynamics.\\
In fact, we find that, out of equilibrium, a particular value of \(\left\langle \bar{z}\right\rangle \) does not always correspond to a unique value of \(\langle \nu \rangle \) , which is the signature of hysteresis. Figure \ref{Fig:Hysteresis}A shows an example where a trait evolves starting at the point \textit{a}, following an optimum that increases towards the point \textit{b.} After equilibrium has been reached, the optimum then moves back to the original value \textit{a}; the other parameters are left unchanged. The trajectories in the space \(\left\langle \bar{z}\right\rangle  ,\langle \nu \rangle  \) are \textit{not} the same for the forward and backwards processes. Although the predictions and the simulations do not agree (clearly, we are in the range where the adaptive peaks are well separated), both, the actual model and the SM approximation show an equivalent qualitative response. In fig. \ref{Fig:Hysteresis}B the optimum moves slowly (as in fig. \ref{Fig:MovingOptimum100lociShortRange} ) and afterwards, moves back to its original point. Because the optimum follows the same trajectory forward and backwards, the amount of hysteresis lower. Finally, in fig. \ref{Fig:Hysteresis}C both the optimum and selection against the genetic variance change abruptly (as in fig. \ref{Fig:SelectionGeneticVariance} ), and after equilibrium is attained, they abruptly change back to their original values. The rates of change of $\langle $\(\bar{z}\)$\rangle $ and \(\langle \nu \rangle \)  have different time scales, and hysteresis is enormous.\\
The puzzling issue here is, why after averaging over the microstates, do the trajectories of the expectations show hysteresis? The question seems be difficult to answer in detailed, mechanistic terms. But, we can understand it from the macroscopic phenomenology. As we showed and discussed above, the trait mean and the genetic variance can evolve more or less independently. Depending on the strength of the selection coefficients, which are the directional component towards the optimum -$N\!\beta $- and the strength of selection again the genetic variance -$N\!\sigma $-, the response will be quicker for one component or for the other (depending which one is effectively bigger). Thus the macroscopic that changes faster in one direction, will be also the one that changes faster in the other direction. For example, in Fig (9A) the optimum shifts. The standing genetic variance initially increases as the rare mutants become frequent and the trait mean increases. Eventually, this variance is rapidly consumed as the trait mean slows down relaxing to equilibrium. If we now consider the reverse process, shifting the optimum back to its original position, the story for the variance will be exactly the same: initial mutants will increase their representation, and the genetic variance increases. But although the genetic variance is taking the same path, the trait mean takes a mirror path: before it was characterized by a period of quick increase, followed by a slow increase. Now it takes a quick decrease followed by a slow decrease.

\begin{figure}[b]
\begin{center}
\caption{Hysteresis on the evolutionary dynamics of a trait that consists of  $n$=100 loci of equal effect. (A) Abrupt
change in the optimum ($N\!\beta $=$\pm $12, $N\!\sigma $=-15, $N\!\lambda $=-1, $N\!\mu $=0.7). (B) Moving optimum
($N\!\beta $ =$\pm $10, $N\!\lambda $ = -1, $N\!\sigma $ = -4, $N\!\mu $ = 0.5; as in fig. \ref{Fig:OptimumAbruptChange100loci}). (C) Abrupt change
in the optimum and selection against the genetic variance ($N\!\beta $=$\pm $12, $N\!\sigma $=-1 $\to $ -5, $N\!\lambda $
=-1, $N\!\mu $ = 0.3 ; as in fig. \ref{Fig:SelectionGeneticVariance}). The arrows indicate the direction of evolution, which start at the points \textit{ s}, ending at
the points \textit{ e}, and then backwards, that is with the forces that changed are switched back. The black lines, are the SM predictions, and
the gray lines, averages from the simulations (population size $N=100$; averages over 500 replicas). In all cases, the paths on both directions
do not overlap, showing that the macroscopic states depend on the path. }
\label{Fig:Hysteresis}
\end{center}
\end{figure}

\includegraphics[scale=0.5]{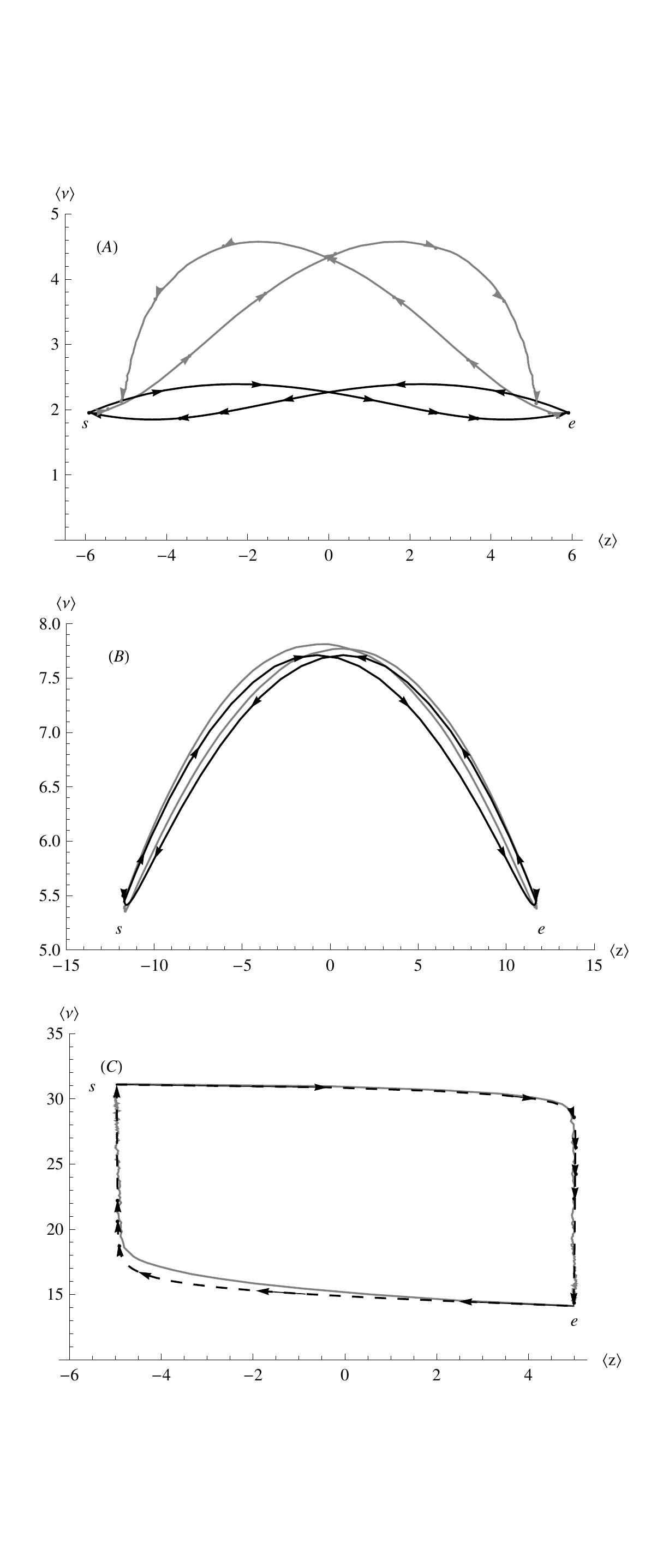}

\section{DISCUSSION}

The main question that we have addressed in this work is whether the SM approximation is valid in the challenging case of stabilizing selection. In general, the maximum entropy distribution has been shown to match that of the exact model (the diffusion equation) at equilibrium, and it is straightforward to determine which macroscopics are --in principle-- relevant for evolutionary change (Appendix 2 in Barton and de Vladar 2009). This extension of our previous methodology has been applied to an important and classical theme in population and quantitative genetics: understanding the factors that generate quantitative variation and its evolution. In the introduction, we posed the problem which refers to the difficulties that appear in estimating the change of a complex trait. We also proposed three ways around to these complications, namely unequal allelic effects, averaging over the different deterministic paths, and genetic drift. We have taken only the last two, for clarity, and assume equal effects to give a worst case scenario where microscopic irregularities are enhanced. In fact, an extension of the SM method to unequal effects involves a slightly different version of the generating function and per-locus equations (see Appendixes 1-2). The approximations follow from the application of the central limit theorem, which does \textit{not} require the same distribution at all loci, only independence between these loci (see Appendix 1).\\
We have been able to predict long term changes of the expectations of the mean and variance of a trait, in the extreme situations when selection either changes abruptly, or gradually due to a moving optimum. In each case the expected trait mean follows radically different paths, yet the expected genetic variance stays roughly constant. Unless there are changes in the selection against the genetic variance, its expectation seldom changes appreciably. Therefore, classical approaches of quantitative genetics, like the breeder's equation (Lande 1979; Falconer 1981, ch. 20), where the constancy of the additive genetic variance is assumed, seem to describe fairly well the evolution of the trait mean, since the fluctuations due to drift do not seem to affect critically this constancy of the expected genetic variance.\\
In our studies of the dynamics, the effective forces other than selection towards the optimum ($N\!\beta $) are weak. This reflects the general principle that selection on a polygenic trait is much more effective on the mean trait than on the variance, by a factor equal to the number of loci. In turn, this is exposed by the local variables, which remain very similar to the actual forces, i.e. \(N\!\sigma^*\sim N\!\sigma\), and \(N\!\mu^*\sim N\!\mu\), and may be related to several factors. First, if the range in which the optimum is changing is narrow (compared with the whole range of response of the trait, $\pm $$n$ , as most traits seem to be), the alleles are unlikely to fix. Even if they have frequencies close to the borders (e.g. if selection against the variance is strong), the variability that is generated by mutation is not reduced through fixation (because we assume $N\!\mu >$1/4). Second, because mutations are frequent, the jumps between peaks provide a mechanism for the change of allele frequencies at an appreciable rate. This is a mechanism which essentially, maintains the genetic variance. Thus, selection on the trait mean can act without affecting the mechanisms that maintain variance. Selection can bias the rates between the peaks in one or the other direction, affecting the trait mean. But again, this does not affect the heterozygosity of the population. This allows the character to be displaced without changing the genetic variance.\\
However, if we select the trait over a wide range, say starting and ending at extreme trait values, we would force fixation of some of the alleles and observe a reduction in the genetic variance toward the extremes, but with a transient inflation in-between. In this case, most mutants appearing in the initial environment are rare, and therefore the variance is small. When selection changes, these mutants become beneficial, and therefore their frequency increases (as has been experimentally documented, see Hoffmann and Meril{\" a} 1999). Eventually, when we approach the other extreme, the $\texttt{'}$good$\texttt{'}$ alleles fix, reducing the variance, as shown in fig. \ref{Fig:MovingOptimum20loci} . \\
Although we have shown that the SM approximation works well in a range of situations, it fails is when the adaptive peaks are well separated (that is when $N\!\mu <$1/4 and/or selection against the genetic variance is strong, \(4|N\!\sigma| \gamma ^2\)$>\!>$1; see fig. \ref{Fig:MarginalDistributions} B). When the distribution becomes peaked close to or at the boundaries, the alleles quickly approach fixation. In this case some of the mutants close to an adaptive peak are forced to jump to a beneficial peak. This process seems to be much quicker than the diffusion times, and therefore the SM predicts a slower response of the genetic variance. The predictions for the trait mean, however, remain accurate. But if the population remains in a regime where there is significant intermediate frequency, the SM method works well.

To summarize, in the light of our theory, we attribute the constancy of the genetic variance to three factors. First, weak selection on the genetic variance. The effective forces on the genetic variance show weak changes, because although the new beneficial phenotypes require a reconfiguration of the genetic states, these can be attained by changing only the proportion of alleles at different loci which are at the different adaptive peaks, rather than a shift on the position of these peaks. Second, the presence of genetic drift. Here, we should also note that although the expectations smoothly change, an individual trajectory will nevertheless fluctuate abruptly. The bounds presented in the above figures indicate the range of these fluctuations. In some occasions, even when the changes in the trait mean or genetic variance are large, these bounds tend to be quite narrow. Therefore even in individual experiments we would expect a regular (i.e. not erratic) pattern of change in the trait mean and genetic variance (seen in reality, e.g. Barton and Keightley 2002, fig. 2). Third, the number of loci composing a trait increases the genetic load, and thus diminishes the rate of change of the expectations of the genetic variance (assuming that selection is on the trait mean). Although this effect is notable in the expectations (figs. \ref{Fig:OptimumAbruptChange5loci} and \ref{Fig:OptimumAbruptChange100loci} ), it seems to be less critical than the above two, as indicated in our theoretical experiments where selection on the genetic variance induces notable changes (fig. \ref{Fig:SelectionGeneticVariance} ).

The above results are in sharp contrast to a $\texttt{'}$general$\texttt{'}$ situation where changes in the genetic variance are not smooth. The implication is that drift is a mechanism that regularizes the course taken by an ensemble of populations through the rugged fitness landscape. Thus, averaging over these drift effects helps us to understand the evolution of quantitative traits. Yet, there is no microscopic hysteresis in our results. We found, curiously, that hysteresis is present in the macroscopic trajectories. This phenomenon results from the combination of two factors. The first is that the genetic variance and the trait mean can evolve independently, each with their own characteristic time scale. The second is that the trajectory of the expectation of the genetic variance is symmetric in time, while that of the trait mean is antisymmetric, meaning that in the backwards dynamics, the trajectory is a mirror image of the forward dynamics. Together, these two factors show not only that genetic systems have memory, but that to $\texttt{'}$degrade$\texttt{'}$ such memory, selection in the contrary direction might be required. Otherwise, the system may get stuck into local optima, especially for low mutation rates, where hysteresis is expected to be enhanced. This last statement is so far hypothesis, which needs to be verified with further analyses.

\section{CONCLUSIONS AND PERSPECTIVES FOR FUTURE RESEARCH}

Our approximation is a step forward in the understanding of the evolution of polygenic traits: it allows us to unravel features using only a few macroscopic measures. Still, traits rarely evolve independently from others. In this sense, there are two plausible extensions to our present work. The first is the explicit coevolution of two or more traits. As in the univariate case, the rate of change of a trait is proportional to the genetic covariances, which are subsumed in the $\mathcal{G}$-matrix  (Lande 1979). The natural question is then how quickly can this matrix respond to selection, mutation, and drift (Steppan et al 2002; Blows 2007). Our method can address this question; it is straightforward to formulate the problem by defining a vector of traits $\langle $\pmb{ \textit{z}}$\rangle $. This would then define a multivariate trait version of the matrices $\mathcal{B}$ and \(\mathcal{C}\), from which we can in principle forecast the evolutionary dynamics responding to simultaneous selection over the different traits. A related problem that we envision is that of pleiotropic effects. Changes in the value of a trait are often accompanied by a change in the fitness of the individual, not only because of its direct effects, but also because many other traits are indirectly being affected (Turelli 1985; Barton 1990; Wagner et al 2008). In this situation, the net effect of the mutations tends to be deleterious. This phenomenon can be quantified by a variable that accounts for all these pleiotropic factors through their net effects on fitness (Turelli 1985; Barton 1990; Keightley and Hill 1990; Kondrashov and Turelli 1992; Gavrilets and De Jong 1993; Orr 2000). \\
Population genetics still faces the challenge of explaining interactions among different factors that have uncertain consequences for the evolutionary process. The SM method, although relying on a moderately complicated mathematical machinery, allows us to systematize several of these factors and to study their effects in a relatively easy way. It is an ideal companion to diffusion methods, and although developed in part as an analogy with SM in physics (Barton and de Vladar 2009; Barton and Coe 2009), it has the same ultimate goals as population and quantitative genetics: to inquire on the factors that produce and maintain variability in genetically complex traits.

\begin{acknowledgements}
The authors would like to thank J. Polechova and F. Palero for comments and discussions.  
This project was supported by the ERC-2009-AdG Grant for project 250152 SELECTIONINFORMATION.
\end{acknowledgements}

\appendix{Gaussian approximation for the generating function}

Following Wright (1937), we can partition the distribution of allele frequencies, and thus the integrand of $\mathbb{Z}$, into the elements of \(\bar{W}^{2N}\) that are associated with the trait mean, and associate those elements which are not, namely:
\begin{equation}
\label{Eq:SplittedPartitionFunction}
\mathbb{Z}=\int \exp\left[2N\!\beta \bar{z}- 2N\!\lambda \bar{z}^2 \right]\exp[2N\!\sigma \nu  +2N\!\mu U]\phi  d^n\pmb{p}\text{  }.
\end{equation}

Now, notice that the second exponential term is separable as a product over loci: Exp[2 $N\!\sigma   \nu $ + 2 $N\!\mu $ \textit{U}] $\phi $ =\( \prod _{i=1}^n  \re^{4N\!\sigma \gamma_i^2p  _iq_i} \left(p  _iq_i\right)^{4N\!\mu-1}.\) If the allele frequencies are not too close to fixation, then the last product (properly normalized) can be approximated by a multivariate Gaussian distribution, centered on the expectation of a given set of macroscopics. We choose these to be: \(\bar{z}\), $\nu $, \(m_3\), \(m_4\) and $H$ which are defined by

\begin{eqnarray}
\left\langle m_3 \right\rangle =2\left\langle \sum_{i=0}^n \gamma _i^3 \left(1-2p_i\right)p_i  q_i\right\rangle  \text{(third moment of the trait)}\\
\left\langle m_4\right\rangle =2\left\langle \sum_{i=0}^n \gamma _i^4  \left(1-2p_i\right)^2p_i  q_i \right\rangle  \text{(fourth moment of the trait)}\\
\langle H \rangle =2\left(\left\langle \sum _{i=0}^n \left(p_iq_i\right)^{-1}\right\rangle -4n\right)  \text{(genetic variance of the log - heterozygosity)}
\end{eqnarray}

These are required to compute the matrices $\mathcal{B}$ and \(\mathcal{C}\). The variances and crossed moments follow directly.\\
Before applying the Gaussian approximation, it is convenient to partition the distribution of each locus around each adaptive peak. For instance, calling \(\psi+=\re^{4N\!\sigma \gamma^2p  q} (p  q)^{4N\!\mu-1}/\tilde{\mathbb{Z}}_o\)  the density around the peak $\texttt{'}$+$\texttt{'}$ peak (close to \textit{p}=1), and similarly \(\kappa_-\) for the $\texttt{'}$-$\texttt{'}$ peak (close to \textit{p}=0), where \(\tilde{\mathbb{Z}}_o\) is their normalization constant (see below), then the product over loci can be expressed as a binomial sum over the pair of peaks of each locus
\begin{equation}
\label{Eq:BinomialPeaks}
\prod _{i=1}^n  \re^{4N\!\sigma \gamma_i^2p  _iq_i} \left(p  _iq_i\right)^{4N\!\mu-1}=\mathbb{Z}_o\prod _{i=1}^n  \left(\right)=\mathbb{Z}_o\sum _{\Bbbk  \epsilon  \mathbb{K}} \left(\prod _{i  \epsilon   \Bbbk , j  \epsilon   \Bbbk ^{c }} \kappa_{i-} \kappa_{j+}\right)\text{  }.
\end{equation}

The index $\Bbbk $ runs in the all the combinations of loci at each peak. For example all loci at the $\texttt{'}$-$\texttt{'}$ peak wold give $\Bbbk $=$\{$1,2,$\ldots $,$n$ $\}$, and its complement would be \(\Bbbk ^c=\{\}\), an empty set. If loci 1,2,4,5 are in the - peak and the rest in the + peak, then $\Bbbk $=$\{$1,2,4,5$\}$ and \(\Bbbk ^c=\{3,6,7,8,\ldots ,\text{\textit{$n$}}\}\). Thus $\mathbb{K}$ contains all these possible permutations. Now we apply the Gaussian approximation (central limit theorem) to the product \(\prod _{i  \epsilon   \Bbbk , j  \epsilon   \Bbbk ^{c }} \kappa_{i-} \kappa_{j+}\), which leads to
\begin{equation}
\label{Eq:GaussianApproxToTheAdditivePart}
\prod _{i=1}^n  \re^{4N\!\sigma \gamma_i^2p  _iq_i} \left(p  _iq_i\right)^{4N\!\mu-1} \simeq  
\frac{\mathbb{Z}_o }{(2\pi )^{5/2} }\sum _{\Bbbk  \epsilon  \mathbb{K}} \text{det}\left(\mathit{c}_{\Bbbk }\right)^{-1/2}\exp\left[-\frac{1}{2}\left(\pmb{a}-\langle \pmb{a}\rangle _{\Bbbk }\right)\cdot \mathit{c}_{\Bbbk }^{-1}\cdot \left(\pmb{a}-\langle \pmb{a}\rangle _{\Bbbk }\right)^{T}\right] , 
\end{equation}
and here, the vector \pmb{ \textit{a}}= \(\left\{\bar{z},\nu ,m_3,m_4, H\right\}\), \pmb{ \textit{a}}$^T$ is its column transpose, and \(\mathit{c}_{\Bbbk }\) is its covariance matrix (its determinant denoted by \(\text{det}\left(\mathit{c}_{\Bbbk }\right)\)). The subscript '$\Bbbk $' indicates that the expectations and covariances are with respect to the distribution at the adaptive peak of the particular configuration $\Bbbk $, and which is free of selection over the trait. The normalization constant is:
\begin{equation}
\label{Eq:AdditiveParitionFunction}
\mathbb{Z}_o = \prod _{i=1}^n  \int _0^{1/2}\re^{4N\!\sigma \gamma_i^2p  q} (p  q)^{4N\!\mu-1}dp = \prod _{i=1}^n  \tilde{\mathbb{Z}}_o^{(i)} .
\end{equation}

We employed the additive property of the functions \(\bar{z}\), $\nu $ and \textit{U}, to separate the integrals, so that the \(\tilde{\mathbb{Z}}_o^{(i)}\)'s are characteristic functions of \textit{per-locus} systems subject to mutation, drift and selection acting over the genetic variance, and therefore depend on the variables $N\!\sigma $ and $N\!\mu $ only: 
\begin{equation}
\label{Eq:PerLocusPFClosedForm}
\tilde{\mathbb{Z}}_o^i=\sqrt{\pi } 4^{-4 N\!\mu} \Gamma (4 N\!\mu)
  {}_1\tilde{F}_1 \left(4 N\!\mu;4  N\!\mu
  +\frac{1}{2};N\!\sigma \gamma _i^2\right)
\end{equation}

($\Gamma $ is the gamma function, and \({}_1\tilde{F}_1\) is the regularized confluential hypergeometric function; Abramowitz $\&$ Stegun 1972, ch. 13). The other
statistics for the adaptive peaks are analyzed in Appendix 2.
Now, introducing eq. (\ref{Eq:GaussianApproxToTheAdditivePart}) into eq. (\ref{Eq:SplittedPartitionFunction}):
\begin{equation}
\mathbb{Z}=\frac{\mathbb{Z}_o }{(2\pi )^{5/2} }\sum _{\Bbbk  \epsilon  \mathbb{K}} \text{det}\left(\mathit{c}_{\Bbbk }\right)^{-1/2}\int  \exp\left[2N\!\beta
 \bar{z}- 2N\!\lambda \bar{z}^2 \right]\times \exp\left[-\frac{1}{2}\left(\pmb{a}-\langle \pmb{a}\rangle _{\Bbbk }\right)\cdot \mathit{c}_{\Bbbk
}^{-1}\cdot \left(\pmb{a}-\langle \pmb{a}\rangle _{\Bbbk }\right)^T\right]d\pmb{a} ~.
\end{equation}

In this integral we should consider a finite range of integration running from the two extremes of each macroscopic, \(a_{\min }\) to \(a_{\max }\).
But for many loci, the density is concentrated near the expectations, thus performing the integral in the whole range is a good approximation. This
of course is true as long as the expectations are far enough from the borders (see ESM 1). Here all the variables except \(\bar{z}\) integrate to
1, and we get 
\begin{equation}
\label{Eq:PartitionFunctionUnequalEffects}
\mathbb{Z}=\mathbb{Z}_o\sum _{\Bbbk  \epsilon  \mathbb{K}} \exp\left[\frac{2}{Q_{\Bbbk }}\left(N\!\beta^2 V_{\Bbbk }+N\!\beta
 M_{\Bbbk }-N\!\lambda M_{\Bbbk }^2\right)\right]/\sqrt{Q_{\Bbbk }} ~,
\end{equation}
where
\[Q_{\Bbbk }=1-4 N\!\lambda V_{\Bbbk }\]

The expectation of the trait mean at each peak \(M_{\Bbbk }\) and its variance \(V_{\Bbbk } \)are given in Appendix 2.

The expression (\ref{Eq:PartitionFunctionUnequalEffects}) is not easy to deal with because the sum is performed over a vast number of terms (all the possible combinations, which amount
to \(2^n\)). One way to make progress is assuming equal effects. In this case, the sum becomes a binomial (see for example in eq. \ref{Eq:BinomialPeaks} that the
sum will directly be a binomial sum). Many of the combinations are equivalent, because all that matters is the number of loci in one peak, or in
the other. Each combination is weighted by the binomial term, denoted by \(\left(
\begin{array}{c}
 n \\
 m
\end{array}
\right)=\frac{n!}{m! (n-m!)}\). A number \textit{ m} of loci standing in a peak contribute to the expected mean by \(m M_o\), where \(M_o\) is the
expected trait mean of a single locus. The other peak will house \textit{ n-m} loci. However, the mean \(M_o\) at the two peaks have equal value
but opposite signs (by convention we take the value at the $\texttt{'}$+$\texttt{'}$ peak), thus its net contribution to the expectation of the trait
mean is \((m-n) M_o\). Adding the contribution of both adaptive peak with a configuration of (\textit{ n-m,m}), the expectation of the mean is thus
\((2m-n) M_o\). The expected variance, however, is invariant with respect to the $\texttt{'}$+$\texttt{'}$/$\texttt{'}$-$\texttt{'}$ peaks. It only
depends on the deviation with respect to the mean value at a peak, which for both peaks is the same, to which each locus contributes by \(V_o\).
The $\texttt{'}$+$\texttt{'}$ peak contributes by \textit{ m} and the $\texttt{'}$-$\texttt{'}$ peak by \textit{ n-m.} Overall, the total variance
is \(n V_o\). Thus eq. (\ref{Eq:PartitionFunctionUnequalEffects}) simplifies, for equal effects, to: 
\begin{equation}
\label{Eq:PartitionFunctionBinomialEqualEffects}
\mathbb{Z} =\frac{\left(\tilde{ \mathbb{Z}_o} \right)^n}{\sqrt{Q}}\sum _{m=0}^n \left(
\begin{array}{c}
 n \\
 m
\end{array}
\right)\times \exp\left[\frac{2}{Q} \left(N\!\beta^2 n V_o-N\!\beta (n-2m) M_o-N\!\lambda (n-2m) ^2M_o^2\right)\right] .
\end{equation}

Although it is feasible to work with this expression it is convenient to approximate the binomial term as a Gaussian. Since the exponential factor
of the sum is also of quadratic order on \textit{ m}, we can obtain a closed expression for the whole sum. Thus approximating \(\left(
\begin{array}{c}
 n \\
 m
\end{array}
\right)\simeq 2^n(\pi  n)^{-1/2}\exp\left[-2\left.(m-n/2)^2\right/n\right]\) , we find that eq. (\ref{Eq:PartitionFunctionBinomialEqualEffects}) becomes
\begin{equation}
\label{Eq:PartitionFunctionEqualEffectsGaussian}
\mathbb{Z} =\frac{\left(\tilde{2 \mathbb{Z}_o} \right)^n}{\sqrt{\pi  n Q}}\sum _{m=0}^n \exp\left[-2\left.(m-n/2)^2\right/n\right]\times
\exp\left[\frac{2}{Q} \left(N\!\beta^2 n V_o-N\!\beta (n-2m) M_o-N\!\lambda (n-2m) ^2M_o^2\right)\right] .
\end{equation}

Now, we replace the sum by an integral over \textit{ m} in the whole real range (-$\infty $,$\infty $), and solve it to get
\begin{equation}
\mathbb{Z} =\frac{\left(\tilde{2\mathbb{Z}_o} \right)^n}{\sqrt{R}}\exp\left[2 N\!\beta^2 n \left.\left(V_o+M_o^2\right)\right/R\right]
\end{equation}
where
\[R=1-4 N\!\lambda n\left(V_o+M_o^2\right) ~.\]

We would like to remind that $N\!\lambda $ is \textit{ always} negative, thus the factors \textit{ Q} and \textit{ R} remains positive for
any choice of parameters.

Although it is tempting to employ the last equation to calculate the macroscopics, this in most cases it will lead to substantial errors. The correct
method is to compute the derivatives from eq. (\ref{Eq:PartitionFunctionEqualEffectsGaussian}) and then perform the Gaussian approximation over the sums of such result.

\appendix{Per-locus statistical mechanics}

\newcommand{\CHypG}[3]{\ensuremath{ {}_1\tilde{F}_1\left( #1; #2; #3 \right)}}
\newcommand{\FHG}{\ensuremath{{}_1\tilde{F}_1}}
\newcommand{\pd}{\partial}

The expectations for the macroscopic variables above, are expressed in terms of the contributions by the different permutations of the loci around each adaptive peak, which we expressed as a Gaussian function. These contributions are also described with a statistical mechanical approximation, but for a polygenic system where selection acts only over the genetic variance, and where mutation and drift affect the population's genetic structure. This system has its own generating function, given by eqns. (\ref{Eq:AdditiveParitionFunction}) and (\ref{Eq:PerLocusPFClosedForm}), from which the expectations and covariances are calculated. The first two expectations that we require for the full system of stabilizing selection are those of the trait mean and its variance at the adaptive peaks, that is \(M_{\Bbbk }\) and \(V_{\Bbbk }\), which are given by:
\begin{equation}
\label{Eq:TraitMeanUneqEffectsNoSelection}
M_{\Bbbk } \equiv \left\langle \bar{z}\right\rangle _{\Bbbk } = \sum _{m=0}^n \left(2 \mathbf{1}_{\Bbbk (m)}-1\right) \frac{\CHypG{4N\!\mu}{4N\!\mu+1}{N\!\sigma \gamma _m^2}}{\CHypG{4N\!\mu}{4N\!\mu+1/2}{N\!\sigma \gamma _m^2}}
\end{equation}
where \(\mathbf{1}_{\Bbbk (m)}\) is an indicator function, which gives 1 if the locus m belongs to $\Bbbk$ and 0 otherwise. The variance is  \(V_{\Bbbk } \equiv text{Var}_{\Bbbk } \bar{z} =\left\langle \bar{z}^2\right\rangle_{\Bbbk
   }-\left\langle \bar{z}\right\rangle _{\Bbbk }^2\)
\begin{equation}
\label{Eq:SquaredTraitMeanUneqEffectsNoSelection}
\left\langle \bar{z}^2\right\rangle_{\Bbbk}=\sum _{m=0}^n \frac{\gamma _m^2}{2} \frac{\CHypG{4N\!\mu}{4N\!\mu+3/2}{N\!\sigma \gamma _m^2}}{\CHypG{4N\!\mu}{4N\!\mu+1/2}{N\!\sigma \gamma _m^2}}
\end{equation}

Similarly, other quantities can be computed from the eqns. (\ref{Eq:AdditiveParitionFunction}) and (\ref{Eq:PerLocusPFClosedForm} by direct integration or by taking derivatives w.r.t. $N/sigma$, to obtain the expectation of the genetic variance for a given configuration across the peaks, or w.r.t. $N\!\mu$  to obtain the expectation of the genetic variability, $U$. And similarly, the second derivatives to compute the variances and covariances. With this method we can obtain the expressions for all the macroscopic variables needed to compute the dynamics of the fully coupled polygenic system.
Under the assumption of equal effects, all the adaptive peaks are equivalent. Thus the product of the per-locus generating functions (eq.\ref{Eq:AdditiveParitionFunction}) reduces to
\begin{equation}
\mathbb{Z}_o = \tilde{\mathbb{Z}}_o^n
\end{equation}
and consequently all the macroscopic variables simplify. Under equal effects,  eqns. (\ref{Eq:TraitMeanUneqEffectsNoSelection}) and (\ref{Eq:SquaredTraitMeanUneqEffectsNoSelection}) become
\begin{eqnarray}
M_o\equiv \left\langle \bar{z}\right\rangle _o = \frac{\gamma  (n-2 m)}{\sqrt{\pi }}  \frac{\CHypG{4N\!\mu}{4N\!\mu+1}{N\!\sigma \gamma ^2}}{\CHypG{4N\!\mu}{4N\!\mu+1/2}{N\!\sigma \gamma ^2}} \\
\left\langle \bar{z}^2\right\rangle_o = \frac{n \gamma ^2}{2}  \frac{\CHypG{4N\!\mu}{4N\!\mu+3/2}{N\!\sigma \gamma ^2}}{\CHypG{4N\!\mu}{4N\!\mu+1/2}{N\!\sigma \gamma ^2}}~.
\end{eqnarray}

Using the generating function, the genetic variance follows directly:
\begin{equation}
\langle \nu \rangle_o = \frac{\pd \log(\mathbb{Z}_o)}{2 \pd N\!\sigma} =2 n N\!\mu \gamma ^2
\end{equation}

Similarly, the expectation of the genetic variability results in
\[\langle U \rangle_o = \frac{\pd \log(\mathbb{Z}_o)}{2 \pd N\!\mu} = 2 n\left(\Psi ^(0)[4 N\!\mu]-\log (4)\right) + \]
\begin{equation}
n \frac{\FHG^{(0,1,0)}\left(4N\!\mu ; 4N\!\mu+1/2 ; N\!\sigma \gamma^2 \right)   +\FHG^{(1,0,0)}\left( 4N\!\mu ; 4N\!\mu+1/2 ; N\!\sigma \gamma^2 \right)   }{\CHypG{4N\!\mu}{4N\!\mu+1/2}{N\!\sigma \gamma _m^2}}
\end{equation}

The function $\Psi ^(k)$ is the polygamma function (k'th derivative of the log-gamma function), and this as well as the derivatives of the hypergeometric functions lack a closed form, and are special functions themselves. Fortunately, in most cases these are readily implemented in numerical packages. See however Abramowitz and Stegun (1972, Chs. 6 and 13) for their definitions in integral or series forms, and their properties.
Other terms of higher order like the variances and covariances follow similarly. However, they involve complicated and long polynomial forms on the hypergeometrics, and writing them down is not particularly useful. However, all the calculations performed in this research are implemented in a Mathematica (v7.0) package supplied in the ESM 2, which includes a brief tutorial (ESM 3).

\nocite{Barton:1989,Barton:1986,Bulmer:1972,TurelliBarton:1990,Rogers:2003p18,MaynardSmith:1979p4690,Kondrashov:1992p2926,Fowler:2002p4625,Kopp:2009p5240,Barton:1987p4186,Hoffmann:1999p4626,Barton:1990p20,Sella:2005p1584,Wright:1937p5322,Barton:2008p226,Turelli:1994p784,Barton:1987p4601,Barton:2009p952,Keightley:1990p5313,Wright:1935p4687,PrugelBennett:1994p3284,Willensdorfer:2003p180,Jones:2004p177,Zhang:2010p4717,Burger:1993p3207,Aita:2005p3280,Wagner:2008,Hill:1982p4691,Barton:1990p4381,Rattray:2001p19,Wright:1935p1872,PrugelBennett:1997p202,Gavrilets:1993,Burger:1991p2347,Hermisson:2005p1495,Wright:1931,Kopp:2007p4362,Aita:2004p3279,Barton:2002p26,Rouhani:1987p4185,Kimura:1962p4335,Rattray:1995p4719,Blows:2007p1949,Roff:1987p4675,Turelli:1985p4718,Zhang:2005p789,PrugelBennett:1997p3209,Orr:2000p3218,Barton:1989p13,Lande:1979p4661,Burger:2005p5278,Sato:2008p5295,Rogers:2000p204,Kimura:1965p1857,Roff:1999p4672,Burger:1995p4394,Steppan:2002p153,Orr:2001p1494,Shannon:1948p4599,Iwasa:1988p12,Waxman:2005p5296,CoxHinkley:1974,CrowKimura1970,Ewens:1979,Falconer:1981,Gardiner:2004,LynchWalsh:1998,AbramowitzStegun:1972,Turelli:1988p4676}


\label{lastpage}
\end{document}